\documentclass[a4paper,11pt]{article}
\pdfoutput=1 % if your are submitting a pdflatex (i.e. if you have
\usepackage{jcappub} % for details on the use of the package, please
\newcommand{\etal}{ {\it et al.}}
\newcommand{\be}{\begin{equation}}
\newcommand{\ee}{\end{equation}}
\newcommand{\bea}{\begin{eqnarray}}    
\newcommand{\eea}{\end{eqnarray}}     

\title{Spatial density fluctuations and selection effects in galaxy
  redshift surveys}

\author[a,b,c,1]{Francesco Sylos Labini} 
\author[d,2]{Daniil  Tekhanovich}
\author[e,3]{Yurij V. Baryshev} 

\affiliation[a]{Centro Studi e
  Ricerche Enrico Fermi, Via Panisperna 89 A, Compendio del Viminale,
  00184 Rome} 

\affiliation[b]{Italy and Istituto dei Sistemi Complessi
  CNR, Via dei Taurini 19, 00185 Rome, Italy}

\affiliation[c]{INFN, Unit of Rome 1, Physics Department, University
  of Rome ``Sapienza'', P.le A. Moro 2, 00185 Rome, Italy}

\affiliation[d]{Faculty of Mathematics and Mechanics, Saint
  Petersburg State University, Staryj Peterhoff, 198504,
  St.Petersburg, Russia}

\affiliation[e]{Institute of Astronomy, St.Petersburg 
State University, Staryj Peterhoff, 198504,
St.Petersburg, Russia} 

 \emailAdd{Francesco.SylosLabini@roma1.infn.it}

 \emailAdd{d.tekhanovich@spbu.ru}

 \emailAdd{y.baryshev@spbu.ru}

\abstract{ One of the main problems of observational cosmology is to
  determine the range in which a reliable measurement of galaxy
  correlations is possible. This corresponds to determine the shape of
  the correlation function, its possible evolution with redshift and
  the size and amplitude of large scale structures. Different
  selection effects, inevitably entering in any observation, introduce
  important constraints in the measurement of correlations. In the
  context of galaxy redshift surveys selection effects can be caused
  by observational techniques and strategies and by implicit
  assumptions used in the data analysis. Generally all these effects
  are taken into account by using pair-counting algorithms to measure
  two-point correlations. We review these methods stressing that they
  are based on the a-priori assumption that galaxy distribution is
  spatially homogeneous inside a given sample. We show that, when this
  assumption is not satisfied by the data, results of the correlation
  analysis are affected by finite size effects.In order to quantify
  these effects, we introduce a new method based on the computation of
  the gradient of galaxy counts along tiny cylinders.  We show, by
  using artificial homogeneous and inhomogeneous point distributions,
  that this method is to identify redshift dependent selection effects
  and to disentangle them from the presence of large scale density
  fluctuations.  We then apply this new method to several redshift
  catalogs and we find evidences that galaxy distribution, in those
  samples where selection effects are small enough, is characterized
  by power-law correlations with exponent $\gamma=0.9$ up to $20$
  Mpc/h followed by a change of slope that, in the range [20,100]
  Mpc/h, corresponds to a power-law exponent $\gamma=0.25$. Whether a
  crossover to spatial unformity occurs at $\sim 100$ Mpc/h cannot be
  clarified by the present data.}  

\begin{document} 
\maketitle
\flushbottom

%\maketitle  IS IGNORED %%%%%%%%%%%

\section{Introduction}

Galaxy redshift surveys represent one of the cornerstone of modern
cosmology. In the past decades we have assisted to an exponential
growth of the data \cite{york2000,colless01,Drinkwater2010,boss} which
have revealed that galaxies are organised in a large scale network of
filaments and voids.  
{ Statistical analysis of these surveys have shown that the galaxy
  distribution is characterised by power-law correlations in the range
  of scales [0.1-20] Mpc/h\footnote{ This situation corresponds to a
    power-law decay of the average conditional density (see below) in
    the range of scales [0.1-20] Mpc/h. Instead, the standard
    two-point correlation function $\xi(r)$ exhibits a break from a
    power law behaviour at about $10$ Mpc/h (for more details see
    \cite{book} and references therein).}: whether or not on scales
  $r> 80$ Mpc/h correlations decay and the distribution crossovers to
  uniformity, is still matter of considerable debate
  \cite{slmp98,Hogg2005,joyce2005,sdss2007,sdss_aea,sdss_epl,2df_epl,2df_aea,gumbel,cqg_review,bagla2007,NabokovBaryshev10a,NabokovBaryshev10b,Einasto2012,Scrimgeour2012,clowes2013,nadathur2013}.
}
This { debate was } originated by the use of different statistical
methods to measure two-point correlations, to estimate statistical and
systematic errors and to control the selection effects that maybe
present in the data { (see, e.g., \cite{slmp98,rees})}.  {Indeed,
  the construction of large enough galaxy redshift catalogues for a
  reliable statistical analysis represents} a complex problem of
observational astronomy.  {In general, observations are} exposed to a
variety of systematic effects that can {non trivially} affect {the}
study of {correlation} properties. For instance, there are effects
which depend on the observational strategy adopted to construct a
particular sample, there are intrinsic physical effects which
ultimately depend on the distance of a galaxy from us (e.g., galaxy
evolutionary corrections, K-corrections, etc.), and corrections that
must be applied to the data to built a three-dimensional sample
suitable for a correlation analysis.  In addition, several corrections
require theoretical modelling {and assumptions} which can bring more
uncertainty {in the results of a data analysis}.  For these reasons it
is important to assess the impact of selection effects on the
statistical information derived from the study of the correlation
function.

{ Standard methods based on pair-counting algorithms may correct for
  selection effects, but they are based on the a-priori assumption of
  spatial uniformity.  In this paper, we introduce an analysis, called
  the gradient cylinder method (GCM), { which is based on a small
    number of a-priori assumptions} and that is suitable to { identify
    selection effects in} any kind of statistically isotropic and
  homogeneous distribution, { bf i.e., even if the distribution is not
    uniform at large scales. This new method allows us} to measure
  unambiguously the presence of redshift dependent selection effects
  in the data and to quantify their impact on the estimations of
  two-point correlations both for spatially homogeneous and
  inhomogeneous point distributions.}  { We stress that the new
  method that we introduce here represents one of the test that can be
  applied to the data to detect radial dependent selection effects and
  it must be intended to be a complementary analysis that can provide
  with further elements to the characterisation of the statistical
  properties of a given sample. As we discuss below it is necessary
  that known luminosity selection effects are taken into
  account. However the usefulness of this new test lies in the fact
  that it can detect radial dependent selection effects that are
  generally  not controllable with standard techniques.}

The paper is organised as follows: firstly, in Sect.\ref{methods}, we
discuss different methods and estimators of two-points
correlations. Then we introduce { the} new method, based on the
computation the gradient of galaxy counts in tiny cylinders. We show
that this method is able to detect redshift-dependent selection
effects in the data. In addition, we show that standard pair-counting
estimators fail to detect the intrinsic correlation properties for
{ intrinsically} inhomogeneous distributions. Then, in
Sect.\ref{toys} we study and calibrate the method by using several
distributions with a-priori known properties.  Results in the
different galaxy samples are then discussed in Sect.\ref{results}.
Finally in Sect.\ref{discussion} we draw our main conclusions.

\section{Statistical methods} 
\label{methods}

Two-point correlations can be characterised by determining the
conditional density \footnote{In order to avoid an { heavy
    notation} we use the symbol $\langle ... \rangle$ for both the
  ensemble and the volume averages.}
\begin{equation}
\label{cond} 
\langle n(r_{12}) \rangle dV_1 dV_2 = \frac{\langle n(\vec{r_1})
    n(\vec{r_2})\rangle}{n_0} dV_1 dV_2 \;,
\end{equation}
where $ n(\vec{r})$ is the microscopic number density and $n_0$ is the
(ensemble) average density of the point distribution. Eq.\ref{cond}
gives the a-priori probability of finding two particles placed in the
infinitesimal volumes $dV_1, dV_2$ around $\vec{r_1}$ and $\vec{r_2}$
with the condition that the origin of the coordinates is occupied by a
particle.
A generic estimator\footnote{ We use the apex $^X$ to specify which 
  estimator we consider: only for the FS estimator --- see below ---
  no apex is used.}  of the conditional density can be written
as
\begin{equation}
\label{class1} 
\left< n^G(r) \right> = \sum_{i=1}^{M}  
\frac{ N_i(r)}{V_i(r)}  \omega_i(r)  = 
\sum_{i=1}^{M} n_i(r) \omega_i(r)  
\;, 
\end{equation}
where $M$ is the number of points in the sample, $N_i(r)$ is the
number of points in the volume $V_i(r)$, { $n_i(r)=N_i(r)/V_i(r)$
  is the density}, and $\omega_i(r)$ is a weight such that
\begin{equation}
 \sum_{i=1}^N \omega_i(r)  = 1 \;. 
\end{equation}
Different estimators correspond to different choices for the weights.
When the volume is exactly the same for all points and it is fully
included in the sample volume then $V_i(r) = V(r) \;$ and $
\omega_i(r) =1 \; \forall i$. This is the so-called full-shell (FS)
estimator that can be simply written
\cite{slmp98,kerscher1999,book,cdm_theo} as
\begin{equation}
\label{estimator_np} 
\left< {n(r)} \right> = \frac{1}{M(r)} \sum_{i=1}^{M(r)}
\frac{N_i(r)}{V(r)} = \frac{1}{M(r)}\sum_{i=1}^{M(r)} n_i(r) \;, 
\end{equation}
where now, because of geometrical constraints, the number of points
that contribute to the the average depends on the scale $r$,
i.e. $M=M(r)$.  It can be shown that this estimator is unbiased, i.e.
its ensemble average {\it in a finite volume} is equal to the ensemble
average.
{ When instead the volume} $V(r)$ may lie partially outside the
sample then the estimator in Eq.\ref{class1} {is biased} and the
amplitude of the systematic bias is, as we discuss in more detail
below, in general unknown \cite{kerscher1999,book,cdm_theo}.

{ Statistical errors affecting the determination of}
Eq.\ref{estimator_np} can be easily determined as
\begin{equation}
  \label{eq:condvar}
  \sigma^2(r) = \frac{1}{M(r)} \sum_{i=1}^{M(r)} n_i^2(r) - \langle
  n(r) \rangle^2 \;. 
\end{equation}
{ At scales comparable with the sample size the situation
  systematic errors become larger than statistical ones
  \cite{sdss_aea}}. { Indeed, on large scales} the problem is that
different determinations of the conditional density $n_i(r)$ are not
independent anymore, as they { explore} the same volume because of
the constraints imposed by the finiteness of the sample\footnote{This
  problem in the literature is known as cosmic variance, but the same
  problem occurs for any kind of signal in a finite sample.}. For this
reason it is not possible to properly estimate { neither the
  conditional density, nor its variance at scales comparable with the
  sample size.}

{ In the literature one may find other methods to estimate
  correlation function errors than that Eq.\ref{eq:condvar}. For
  instance several authors suggest that jack-knife methods actually
  recover the correct variance on large scales, but fail on smaller
  scales \cite{Norberg2008}. However, as it was discussed by the
  jack-knife is not a suitable method to estimate errors in the
  presence of long-range correlations for the lack of independence
  between different sub-samples \cite{fsletal_absence}. Thus we will
  adopt the most conservative one given by Eq.\ref{eq:condvar}. }

Note that the statistical errors computed through Eq.\ref{eq:condvar}
are in general negligible at all scales as long as the average is
performed over a number of centres $M(r) > 10^2$ (as we do it what
follows). The main source of error in the measurement of the
conditional density is { thus} represented by systematic errors,
which are connected to finite-volume effects.  In the past we have
tested different methods { to control and quantify such effects; we
  concluded that the most reliable one is represented by study, for
  each $r$, the full probability density function of the values of
  $n_i(r)$ entering in Eq.\ref{estimator_np} (we refer the interested
  reader to \cite{sdss_aea,2df_aea,gumbel,cqg_review} for a more
  detailed discussion of this issue). }

The most natural and simpler choice for the volume $V(r)$ in the case
of the FS estimator is a sphere. In this case the analysis is limited
by the radius $R_s$ of the largest sphere that can be fully included
in the sample volume. For typical surveys $R_s$ can be much smaller
than the largest separation between { the most distant pairs} of
galaxies $R_{max} \gg R_s$.  A way to avoid this constraint is to
employ the non-FS estimators of the correlation function used
standardly in the cosmological literature
\cite{dp83,rivolo,hamilton,ls,kerscher1999}.  Another possibility that
we study in the next subsection is to implement the FS estimator in
cylinders.  We then review the problems of the non-FS estimators in
Sect.\ref{pair}.

\subsection{The full shell estimator in cylinders} 
\label{full-shell-estimator-in-cylinders}

In what follows, in Eq.\ref{estimator_np} we consider $V(r)$ to be the
volume of a cylinder of height $2r$ and radius $h$ fully included in
the sample volume\footnote{Note that we use $r$ for the cylinder half
  height because a galaxy is place in the middle of the cylinder and
  its distance from the bases $r$ is varied.  Instead, the cylinder
  radius, $h$, is fixed.}. The point $i^{th}$ is now placed in the
centre of the cylinder at distance $r$ from each of the two bases and
in the centre of the circle of radius $h$. The cylinder volume is
simply
%\begin{equation}
%\label{cylvol}
\[
V(r) = 2 r \times \pi h^2 \;. 
\]
%\end{equation}
%
We take the generic cylinder to be oriented in a manner parallel to
the line of sight (LOS) or perpendicularly to it\footnote{Note that in
  the latter case the cylinder is oriented perpendicularly only to the
  LOS that passes for the galaxy chosen as centre of the cylinder
  itself. }.  In the  {former} case we determine the estimator $
\left<{n^p(r;h)}\right>$, while in the {latter} case 
$\left<{n^o(r;h)}\right>$. If the distribution is isotropic we expect
that
\begin{equation}
\label{iso} 
 \left< n^p(r;h) \right> = \left< n^o(r;h) \right> = \left< {n(r)} \right>
\end{equation}
where $\left< n(r) \right>$ is determined in spheres of radius $r$ 
(i.e., Eq.\ref{estimator_np}).

\begin{figure}[tbp]
\centering 
\includegraphics[width=.45\textwidth]{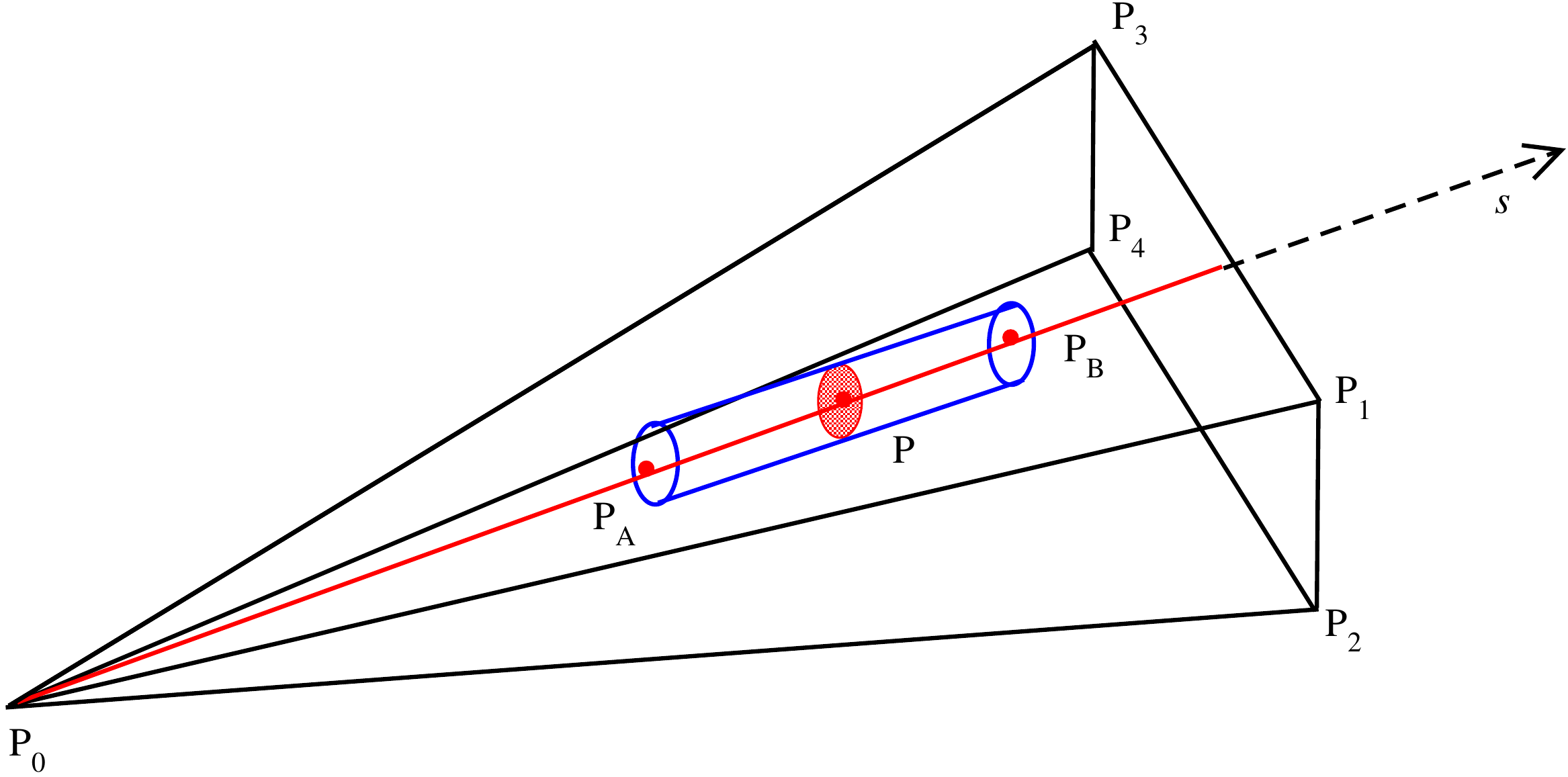}
\includegraphics[width=.45\textwidth]{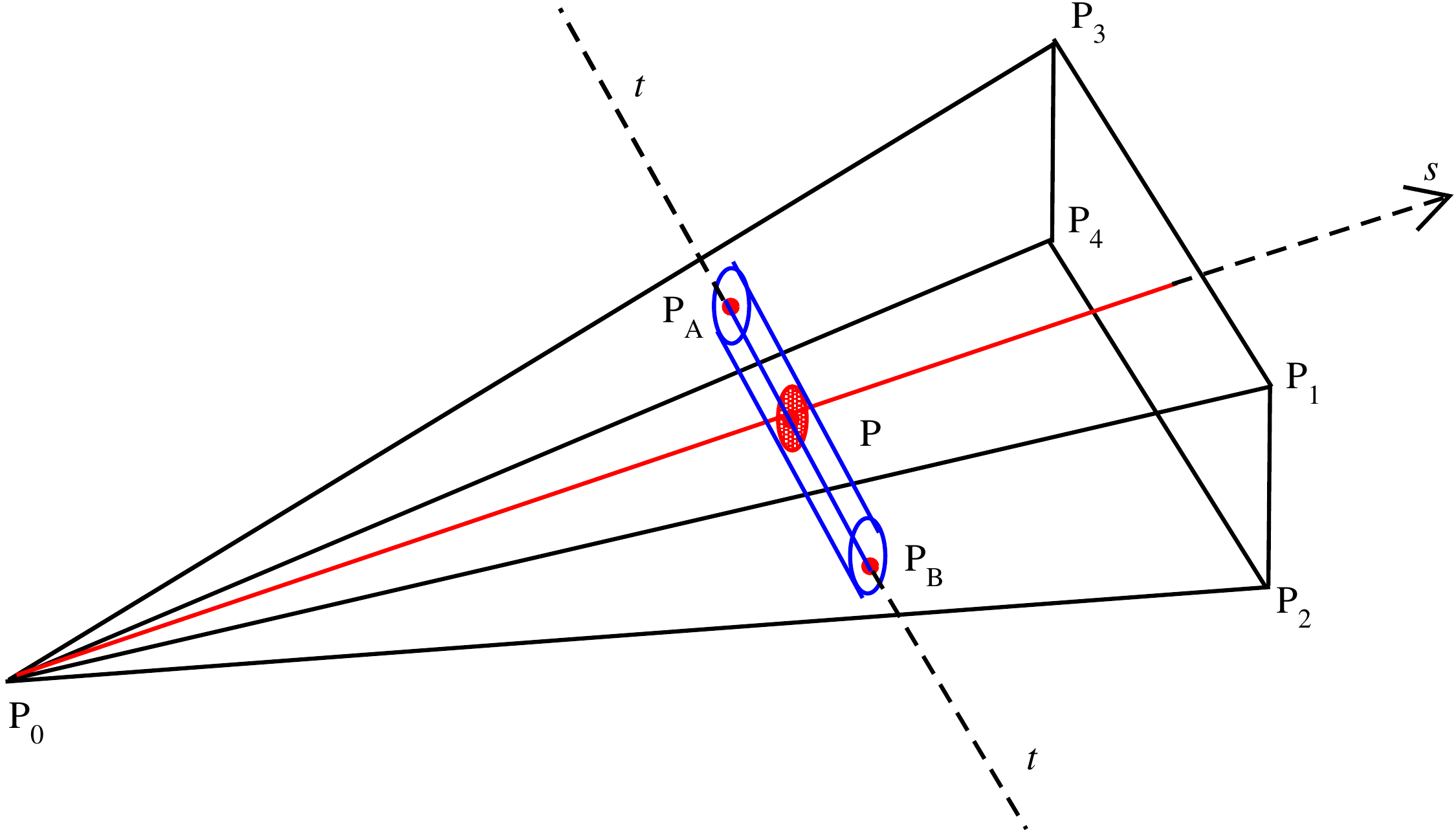}
\hfill
\caption{Left panel: geometry of the cylinder oriented along the line
  of sight.  Right panel: geometry of the cylinder oriented
  orthogonally to the line of sight.\label{geometry1-grad}}
\end{figure}

The determination of whether the $i^{th}$ point can be included or not
in the sum in Eq.\ref{estimator_np} requires the measurement of
several distances necessary to control that the cylinder is fully
included in the sample volume.  In { left} panel of
Fig.\ref{geometry1-grad} its shown the case where the cylinders are
oriented along the LOS and in { right} panel the case in which the
cylinders are oriented perpendicularly to the LOS: the points $P_1
,P_2, P_3, P_4$ determine { the points where the straight lines
  from the origin} of the coordinates $P_0$, intersect the boundaries
of the sample (the generic galaxy is called $P$ and it is placed in
the middle of a cylinder of height $2r$ and radius $h$, whose extremes
are placed in the points $P_A$ and $P_B$).

For cylinders oriented parallel to the LOS, the points $P_A$ and $P_B$
are placed on the line $s$ at distance $\pm r$ from the point $P$
while for cylinders oriented perpendicularly to the LOS the points
$P_A$ and $P_B$ are placed on the line $t$ at distance $\pm r$ from
the point $P$. Note that in the former case it is necessary to choose
an orientation for the line $t$ orthogonal to $s$: for convenience, we
choose parallel to the plane identified by the points $[P_0,P_2,P_4]$,
but any other choice would be equally valid.

In summary, to verify that a generic cylinder is placed inside the
boundaries of the sample, we have to control that the points $P_A$ and
$P_B$ are indeed contained the sample volume and that their distance
from the sample boundaries, orthogonally to the straight line $s$ or
$t$, is smaller than $h$.  
Given these constraints, we compute { for} each cylinder, { with
  the $i^{th}$ galaxy at its centre point, the number of points
  $N_i(r;h)$ contained in it; the density entering in the sum in
  Eq.\ref{estimator_np} , is then}
%\footnote{ In order to avoid a
%  heavy notation, we will denote the density in Eq.\ref{cylvol} as
%  $n_i(r)=n_i(r;h)$, leaving implicit the dependence on the cylinder
%  radius $h$. }
%%
\begin{equation}
\label{cylvol}
 n_i(r;h) = \frac{N_i(r;h)}{2 \pi r h^2} \;.
\end{equation}
{ In what follows $h$ is chosen to be slightly larger than the
  average distance between nearest neighbours: for smaller values of
  $h$ the number of points of contained in a generic cylinder of
  length $r$ is not sufficiently large to avoid a too high Poisson
  noise.}

As mentioned above, in many cases the radius of the largest sphere
contained in the sample volume $R_s$ is much smaller than the maximal
distance between two galaxies $R_{max} \gg R_s$. In particular this
situation occurs when the survey is very deep and covers a small solid
angle in the sky \cite{book}. Thus by measuring the conditional
density in cylinders one can largely increase the range of
measurements with a FS estimator: in principle it is possible to
reach $R_{max}/2$ instead of $R_s$. 
{ However one must consider that this estimation involves the
  convolution of the correlation properties of a distribution with the
  cylinder window function.}  In order to understand the effect of the
  convolution of the cylinder window function with the shape of the
  two-point correlation function let us consider a few simple
  examples. We firstly suppose that in a sphere of radius $r$ the
  average conditional density has a simple power law behaviour:
\begin{equation}
\langle n(r) \rangle = \frac{A}{r^\gamma}
\end{equation}
with $0 \le \gamma <3$.  The density of points in the cylinder is 
\be
\label{general}
\langle n(r; h)\rangle = \frac{2}{2\pi r h^2} \int_0^r dx \int_0^{2\pi}
d\phi \int_0^{h} \frac{A \rho d\rho} {(x^2+\rho^2)^{\gamma/2}} \;.
\ee
For $\gamma=0$ (Poisson distribution)  
{ we simply find} 
$\langle n(r)\rangle = A \;.$
When $\gamma>0$, by numerically integrating Eq.\ref{general}, we find
\be 
  \langle n(r; h)\rangle \approx \zeta (h) r^{3-\gamma}  \;, 
\ee
where the amplitude $\zeta (h)$ depends on $h$ (which is taken fixed)
while the corrections to the leading $r^{3-\gamma}$ behaviour are negligible
when $h \ll r$.

Let us now consider the case in which the conditional density { in
  spheres} has a change of { slope: }
\bea 
\label{twopower} 
\langle n(r) \rangle  = \frac{A}{r^{\gamma_1}} \;\; \mbox{for}
\;\; r\le \lambda_* \\ \nonumber \langle n(r) \rangle =
\frac{B}{r^{\gamma_2}} \;\; \mbox{for} \;\; r\ge \lambda_* \eea
with $0 \le \gamma_2 \le \gamma_1 <3$ and where $B=A
\lambda_*^{\gamma_2-\gamma_1}$ for continuity reasons.  In addition,
for simplicity we consider  $h <
\lambda_*$.

\begin{figure}[tbp]
\centering 
\includegraphics[width=.45\textwidth]{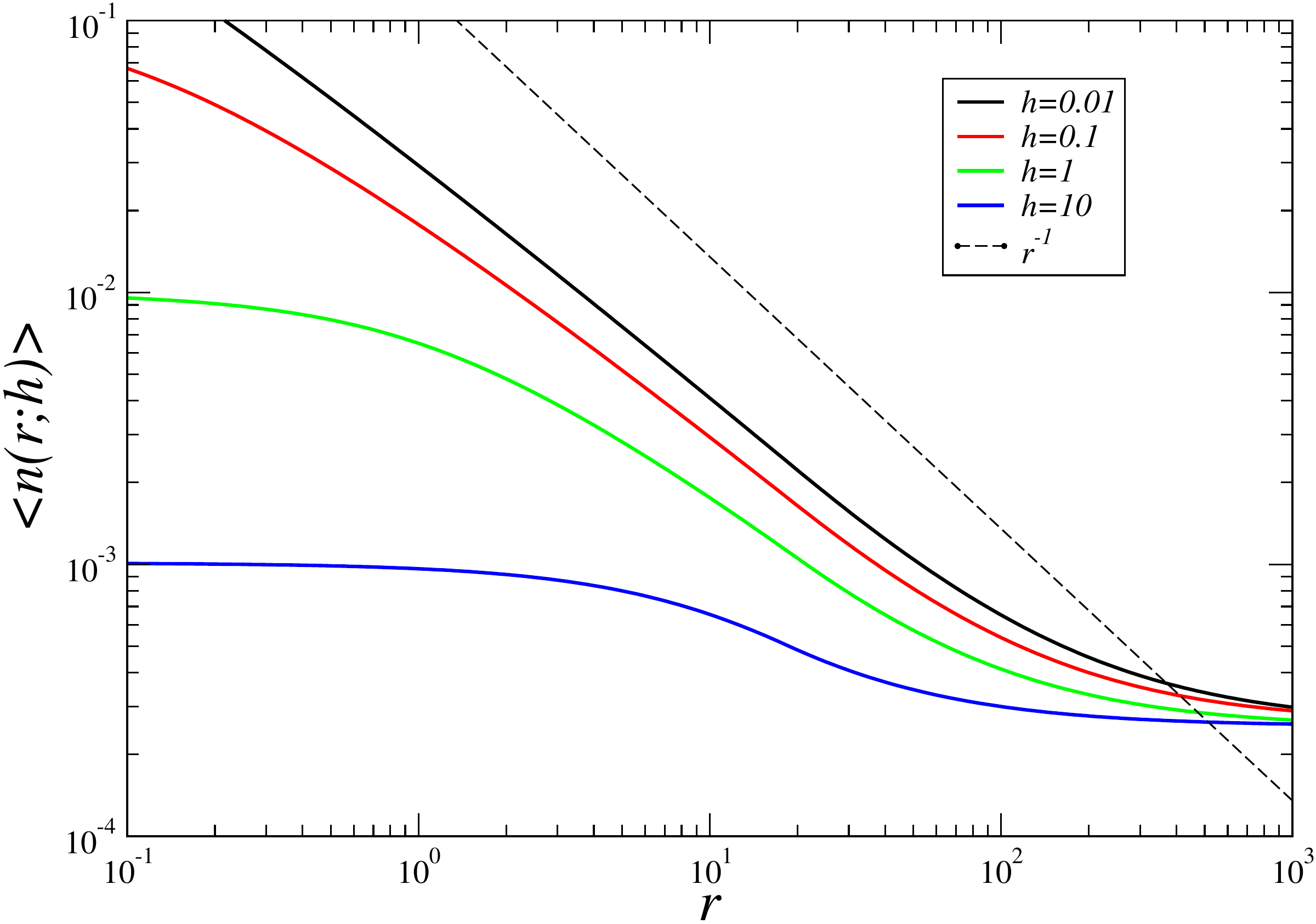}
\includegraphics[width=.45\textwidth]{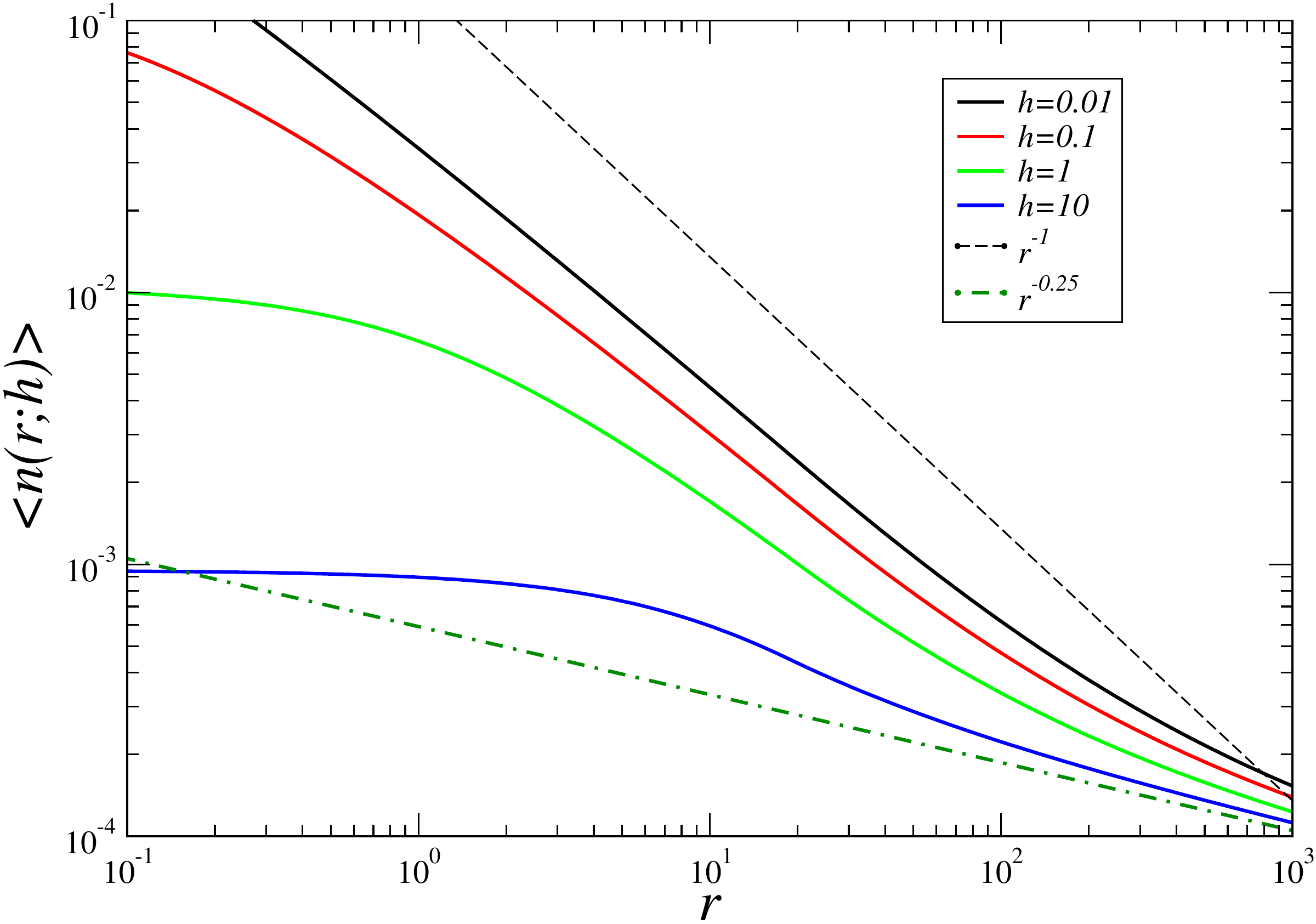}
\hfill
\caption{Behaviour of the conditional density given by
  Eq.\ref{twopower} estimated in cylindrical volumes of radius $h$.
  Left panel: $\gamma_1=1.0; \gamma_2=0.0\; $ and $\lambda_0 =20 $
  Mpc/h.  Right panel: $\gamma_1=1.0; \gamma_2=0.25\; $ and
  $\lambda_0 =20 $ Mpc/h. \label{fig_1.0_0.0_20}}
\end{figure}

The results of the numerical integration for two particular cases
(i.e., $\gamma_1=1$, $\lambda_0=20$ Mpc/h and $\gamma_2=0$ or
$\gamma_2=0.25$) is shown in Fig.\ref{fig_1.0_0.0_20}: { we can
  clearly conclude that,} when the conditional density has a single
power-law behaviour, by computing it in cylinders, one is able to
measure the correlation exponent properly on scales larger than the
cylinder radius $h$ { thus extending the analysis to $R_{max} >
  R_s$}.  Instead, when the conditional density has a change of slope
{ it is not possible, unless the samples extends over several
  decades, to reach a clear conclusion about the large scale
  exponent.}

%%%%%%%%%%%%%%%%%%%%%%%%%%%%%%%%%%%%%%%%%%%%%%%%%%%%%%%%%%%%%%%

\subsection{Pair-counting estimators} 
\label{pair}

Estimators of two-point correlations used standardly in the
cosmological literature \cite{dp83,rivolo,hamilton,ls,kerscher1999},
are based on the determination of $\xi(r)$. This is related to the
conditional density by the equation \cite{book} 
\be 
\langle n(r) \rangle = {\bar {n}} \left(
1+\xi(r) \right)
\label{xi-gamma}
\ee
where ${\bar {n}}$ is the sample average density. It is well known
that { for } inhomogeneous distributions the estimation of the sample
density ${\bar {n}}$ depends explicitly on the sample size { $R_s$ so
  is the estimation of $\xi(r)$ \cite{book} .  On the other hand, for
  { distributions that are spatially uniform at scales smaller than
    the sample size} $\langle n(r) \rangle$ is not affected by the
  sample finiteness and thus there is in principle no problem in the
  use of Eq.\ref{xi-gamma}, { to } jointly determine ${\bar {n}}$
  and $\xi(r)$ { for measuring} $\langle n(r) \rangle$.}
 {The situation is however not so simple. As emphasised in \cite{book}
   a great care must be taken in interpreting the results obtained
   from pair counting estimators from around the scale, of the order
   of $R_s$, at which partial shells begin playing an important role
   in these estimators, i.e. beyond the scale up to which one can
   calculate the FS estimator.  The main problem is that the bias and
   variance of such estimators is calculable only in very simple cases
   (e.g., for the uncorrelated Poisson case), and it is uncontrolled
   { (i.e.  there is not an analytical control of their amplitude)}
   for a broader class of distributions (see { for more details}
   \cite{kerscher1999,kerscher00,cdm_theo}).}
That is, the bias and the variance of these estimators cannot be
computed for a general point distribution and thus one can only
constraint them through the use of a numerical simulations.  {
  Therefore only if assumes which are the large scale properties of a
  distribution one can measure reliable error bars through numerical
  simulations.}

In what follows we will consider three popular (non full shell)
estimators based on pair counting.  The first one is the so-called
Rivolo estimator \cite{rivolo} that gives equal weighting of all
centres i.e. $\omega_i=\frac{1}{N}$ (see Eq.\ref{class1}) and thus
gives equal weight to partial and full shells. For this reason it will
have a variance which increases strongly at scales comparable to
$R_s$.  This can be written as
\begin{equation}
\label{rivolo} 
\frac{\left< n^R(r) \right>}{n_0} 
= \frac{N_R}{N_D(N_D-1)} \sum_{i} \frac{D_i(r)}{R_i(r)}  
\end{equation}
where { ${N_D}$ (${N_R}$) is the number of data (random) points in
  the volume $V$, $n_0=N_D/V$ is the estimation of the sample density,
  and $D_i(r)$ { ($R_i(r)$) } is the number of data (random) points in
  the distance range $(r-\Delta/2, r+\Delta/2)$ from the point $i$.}
  {In what
  follows, we will consider the integrated value of this estimator,
  i.e. $D_i(r)$ and $R_i(r)$ calculated not in shells but in spheres,
  as this version of the Rivolo estimator was used by
  \cite{Scrimgeour2012}.
}

 {A very widely used estimator of $\xi(r)$ is the one introduced by
   { Davis \& Peebles (DP)} \cite{dp83}: the corresponding estimator of
   the conditional density is}
\begin{equation}
\label{dp}
\frac{\left< n^{DP} (r) \right>}{n_0} = \frac{2N_R}{N_D
  -1}\frac{DD(r)}{DR(r)}
\end{equation}
where $DD(r)$ ($DR(r)$) is the number of data-data (data-random) pairs
with separation in the range $(r-\Delta/2, r+\Delta/2)$. It is easy to
verify that this corresponds a choice of weighting in Eq.\ref{class1}
(see \cite{book})
\[
\omega_i(r)= \frac{ \Delta V_i(r) }{ \sum_i \Delta V_i(r) } \;. 
\]
Partial shells are thus weighted in proportion to their volume. The
idea for this choice is that this may compensate for certain
distributions for the additional variance associated to the partial
shells \cite{kerscher1999,book}.

 {The other estimator that  we consider is related to the one 
introduced by { Landy \& Szalay (LS)} \cite{ls}:} 
\begin{equation}
\frac{\left< n^{LS}(r) \right>}{n_0} = \frac{N_R(N_R-1)}{N_D(N_D -1)}
\frac{DD(r)}{RR(r)} -2\frac{N_R-1}{N_D} \frac{ DR(r)}{RR(r)} + 2 \;.
\label{ls}
\end{equation}
 { The LS} estimator is the most popular in the cosmological
 literature because it has the minimal variance for a Poisson
 distribution of points, with a variance which is proportional to
 $1/N$ rather than $1/\sqrt{N}$ for the other estimators
 considered. This estimator was found to have the minimal variance
 also in the case of a { distribution extracted from a standard
   LCDM simulation.}   \cite{kerscher00}.
{ Note that the theoretical errors in the correlation function can
  be estimated analytically for any uniform distribution once it is
  given the power spectrum \cite{feldman1994}: the difficult problem
  lies in the estimation of the errors in the regime of strong
  clustering.}

%%%%%%%%%%%%%%%%%%%%%%%%%%%%%%%%%%%%%%%%%%%%%%%%%%%%%%%%%%%%
%%%%%%%%%%%%%%%%%%%%%%%%%%%%%%%%%%%%%%%%%%%%%%%%%%%%%%%%%%%%
%%%%%%%%%%%%%%%%%%%%%%%%%%%%%%%%%%%%%%%%%%%%%%%%%%%%%%%%%%%%
%%%%%%%%%%%%%%%%%%%%%%%%%%%%%%%%%%%%%%%%%%%%%%%%%%%%%%%%%%%%

\subsection{The gradient of galaxy counts in cylinders} 
\label{GCM}

In addition to the estimation of the conditional density in cylinders
{ we can compute the galaxy counts gradient} in cylinders oriented
along the LOS: we show below its usefulness. This can be estimated by
\be
\label{deltap} 
\delta^p(r;h) = \frac{\langle N_N(r;h) \rangle- \langle N_F(r;h)
  \rangle}{\langle N(r;h) \rangle} \;, \ee
where we have defined 
\be 
\label{avnumbp}
\langle N(r;h) \rangle = \frac{1}{M(r)} \sum_{i=1}^{M(r)} N_i(r;h) \;,
\ee
while $\langle N_N(r;h) \rangle $ ($\langle N_F(r;h) \rangle$) is
defined similarly to Eq.\ref{avnumbp} but it is computed in a cylinder
of length $r$ (instead of $2 r$) and radius $h$. { In particular,
  for} $\langle N_N(r;h) \rangle $ ($\langle N_F(r;h) \rangle $) the
$i^{th}$ centre-point is located at the base which has the largest
(smallest) radial distance (see the upper panel of
Fig.\ref{geometry1-grad}).

Similarly to Eq.\ref{deltap} we define the analogous quantity for
cylinders oriented orthogonally (see the bottom panel of
Fig.\ref{geometry1-grad}):
\be
\label{deltao} 
\delta^o(r;h) = 
\frac{\langle N_L(r;h) \rangle- \langle N_R(r;h) \rangle}
{\langle  N(r;h) \rangle} 
\ee 
where $\langle N_L(r;h) \rangle$ is the average number of points
contained in the half cylinder delimited by $[P,P_A]$ and $\langle
N_R(r;h) \rangle$ in the other half cylinder delimited by $[P,P_B]$.

 {In what follows we will show that the joint determination of
   $\delta^p(r;h)$ and $\delta^o(r;h)$ in a galaxy catalogues, may shed
   light on the presence of intrinsic fluctuations and extrinsic
   radial dependent selection effects. We will refer to these
   determinations as the gradient cylinder method (GCM).}

{ Note that, in general, that $\delta^p(r)$ ($\delta^o(r)$) is
  non-zero when there is a systematic difference between $\langle
  N_N(r;h) \rangle$ and $ \langle N_F(r;h)$ (or between $\langle
  N_L(r;h) \rangle$ and $ \langle N_R(r;h)$) which is persistent in
  space.  On the one hand, when this is due to a selection effect in
  the direction parallel (orthogonal) to the line of sight
  $\delta^p(r)$ ($\delta^o(r)$) will show a well defined trend, i.e. a
  systematic increase or decrease of its amplitude with the scale
  $r$. On the other hand, when this is due to large scale structures
  $\delta^p(r)$ ($\delta^o(r)$) will be characterised by fluctuations
  so that its amplitude will be different from zero for a range of
  scales correspondent to the spatial extension of the structures. }

\section{Tests of selection effects on artificial distribution} 
\label{toys} 

In order to illustrate the usefulness of the GCM introduced in
Sect.\ref{GCM}, we apply it to point distributions with known
properties that { can be, or not, affected} by a radial selection
$f(s)$, again with controlled properties. { The exercises that are
  discussed in what follows} show that the intrinsic correlation
properties can be distorted by a radial selection function $f(s)$ {
  (where $s$ is the radial distance from the Earth)} and that the
study of the quantities $\delta^p(r;h)$ and $\delta^o(r;h)$
(Eqs.\ref{deltap}-\ref{deltao}) can {reveal} the presence of a radial
selection.  { { We show that the} application of the GCM works
  equally well for uniform and inhomogeneous distributions}. { On
  the other hand, to emphasise the limits of standard pair-counting
  estimators, we also consider some example of the determination of
  the conditional density with pair-counting methods. These will be
  useful to show that both the bias and the variance of these
  estimators, determine their large scale behaviour. }

{ We note that the artificial samples that we considered to test
  our method are already volume limited. We assume that the
  observational selection effect related to the limit in apparent
  magnitude imposed by the observations, as it is a well known
  problem, has been already taken into account in the construction of
  a proper sample for statistical analysis.}

{ The artificial galaxy distributions, each with $N=10^5$ points,
  have been generated, in a box of side $L=1000$ Mpc/h, by using three
  different algorithms: (1) a purely Poisson process (2) a random walk
  and a (3) a random trema dust \cite{book}. We have then chosen a
  random point close to the edge of the cube to be the origin of
  coordinates of a spherical volume limited by ascension right $\in
  [0,\pi/2]$, declination $\in [0,\pi/2]$ and radial distance from the
  origin $s \in [0,500]$ Mpc/h. We have then applied to the artificial
  galaxy a radial dependent selection of points described by the
  following function of the (radial) distance from the origin of
  coordinates:
\be
\label{radselfunc} 
f(s; \alpha) \sim s^{-\alpha} \;. 
\ee
The amplitude of the selection function is tuned to eliminate a
relevant fraction (i.e., $> 20 \%$) of the original distribution
points.

The radial density clearly becomes 
\be
\label{radselfunc2} 
n(s;\alpha) = n(s) \times f(s; \alpha) \;. 
\ee

In addition, a mock galaxy distribution was constructed from a
cosmological LCDM N-body simulation in a box of side $l=500$
Mpc/h\cite{croton}. Even in this case we have then randomly chosen a
mock galaxy close to one of the edges of the simulation box to be the
origin of coordinates. We have then cut a spherical volume limited as
for the artificial distributions discussed above. Then we have applied
to these data a selection function again described by
Eq.\ref{radselfunc2}.  }

\subsection{Poisson}

{ We first consider a Poisson distribution: 
 given the absence of spatial correlations } we find
\be
\label{poisson}
n(s)|_{s=r} \approx \left< {n(r)} \right> \approx \left< {n^p(r;h)}
\right> \approx \left< {n^o(r;h)} \right> \approx const.  \;, 
\ee
where $n(s)$ are the galaxy counts per unit volume as a function of
the { radial distance $s$ from us} and where the equality in
Eq.\ref{poisson},{ valid for $r=s$}, is due to the fact that the
  point distribution has no correlations.  In addition we find
  $\delta^p(r;h) \approx \delta^o(r;h) \approx 0$, in agreement with the
  expectation for { an isotropic distribution of points.}

\begin{figure}[tbp]
\centering 
\includegraphics[width=1\textwidth]{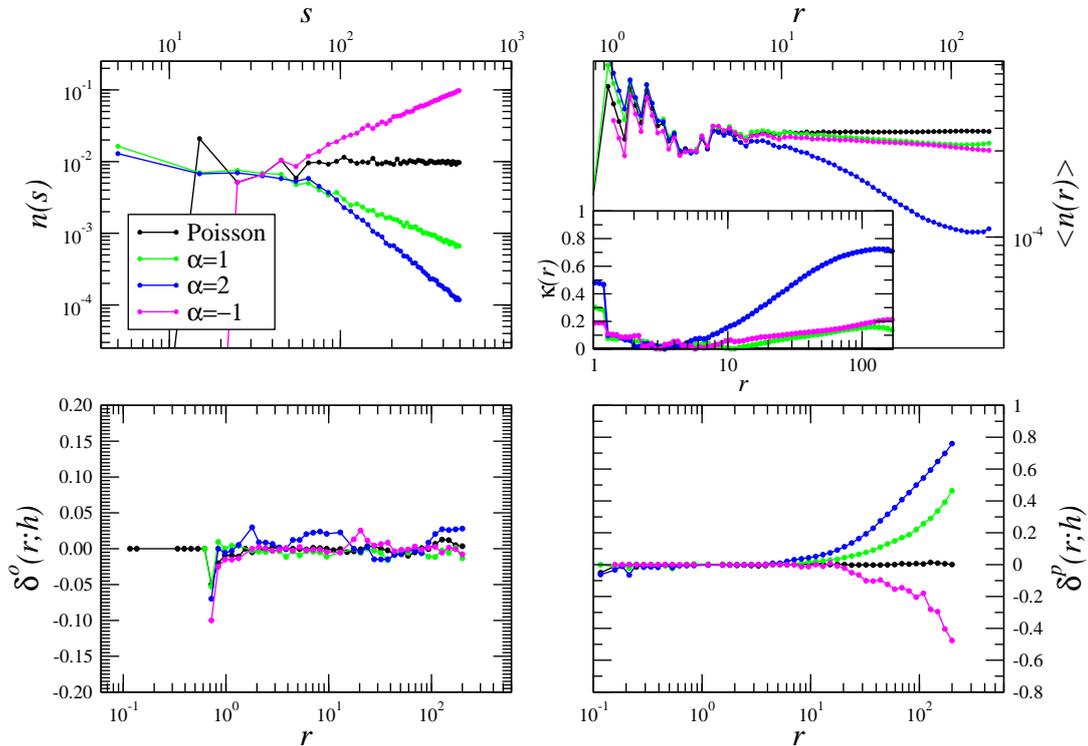}
\hfill
\caption{The case of a Poisson distribution to which it was
          applied a radial selection function described by
          Eq.\ref{radselfunc}: the value of the exponent $\alpha$ is
          reported in the labels. Upper left panel: radial density
          $n(s)$. Upper right panel: conditional density $\left<
          {n(r)^{s}} \right>$ (in the inset panel it is reported the
          behaviour of $\kappa(r)$ --- see Eq.\ref{Kappa}). Bottom left
          panel $\delta^o(r;h=1)$ (Eq.\ref{deltao}).  Bottom right
          panel: $\delta^p(r;h=1)$ (Eq.\ref{deltap}) 
	\label{poisson_selection}}
\end{figure}

{ We have then applied to the data a selection function of the type
described by Eq.\ref{radselfunc2}} 
 {In Fig.\ref{poisson_selection} we report the results obtained by
   changing $\alpha~\in~[-1,2]$}: { in this case we simply find $
 n(s;\alpha) \sim s^{\alpha}$} \footnote{Note that the amplitudes of
   $n(s)$ and $\left< {n(r)} \right>$ for different $\alpha$ have been
   arbitrarily normalised to have the same small scales value.}.

{ In order to measure the departure of the conditional density for
  $\alpha \ne 0$, i.e. $\left< {n(r)} \right>_{\alpha\ne0}$ from its
  unperturbed shape $\left< {n(r)} \right>_{\alpha=0}$ we may the
  quantity} 
\be
\label{Kappa}
\kappa(r) = \left| \frac { \left< {n(r)} \right>_{\alpha\ne0} - \left<
  {n(r)} \right>_{\alpha=0}} { \left< {n(r)} \right>_{\alpha=0}} \right|  
\ee 
{  which measures the percentage change in $\langle n(r) \rangle$
  when a radial selection effect is imposed to the underlying
  distribution. We note that a $\delta^p (r) = 0.1$ is indicative
  approximately of a 10\% change in $\langle n(r) \rangle$.  The
  conditional density has a percentage change in its amplitude only
  when the exponent of the radial selection function is $\alpha \ge
  1$: correspondingly $\delta^p(r;h)$ grows with $r$ as $\langle
  N_N(r;h) \rangle > \langle N_F(r;h) \rangle$ (see
  Eq.\ref{deltap}). Note we find $\delta^o(r;h) \approx 0$ as there is
  not any redshift dependent selection effect in the direction
  orthogonal to the LOS}.

%%%%%%%%%%%%%%%%

\subsection{Strongly correlated distributions} 

\begin{figure}[tbp]
\centering 
\includegraphics[width=1\textwidth]{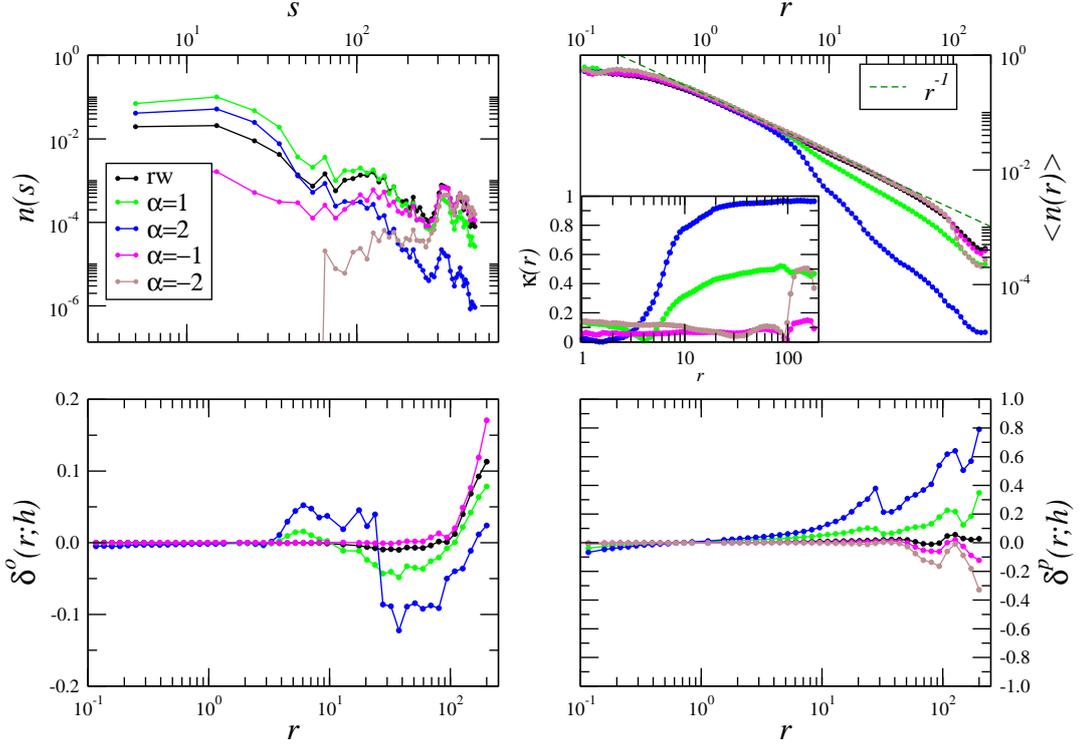}
\hfill
\caption{
As Fig.\ref{poisson_selection} but for case of strongly
  correlated distribution, corresponding to a fractal with $D=2$. 
\label{rw_selection_grad}
}
\end{figure}

\begin{figure}[tbp]
\centering 
\includegraphics[width=1\textwidth]{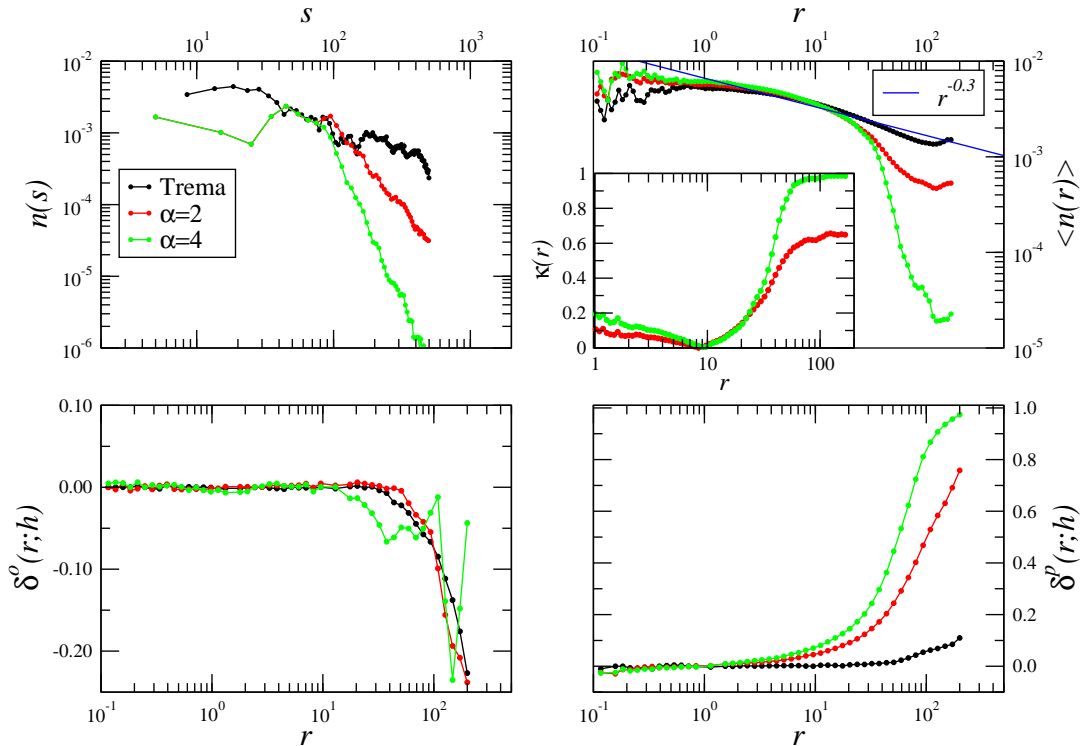}
\hfill
\caption{As Fig.\ref{poisson_selection} but for case of strongly
  correlated distribution, corresponding to a fractal with $D=2.7$.
\label{trema_selection_grad}
}
\end{figure}

 {We have repeated the same tests on strongly correlated
  distributions: (i) a fractal with dimension $D=2$ generated by a
  random walk and (ii) a fractal with $D=2.7$ generated by a trema
  dust algorithm \cite{book} \footnote{ In case of trema dust, we
    distribute $N_c$ random points inside a cube of side $L=500$ Mpc;
    each of them is then considered as centre of a sphere of volume
    $V=(D-3)/n \times 500^3$, where $n=1,..,N_c$, and $D$ is the
    fractal dimension. The fractal set is obtained by distributing
    random points in that part of the cube which does not overlap with
    any of the $N_c$ spheres. In our case, we choose $D=2.7$.}.  As for
  the Poisson distribution we have applied to the data a radial
  selection of the type described by Eq.\ref{radselfunc} with
  different $\alpha$ (see results in Fig.\ref{rw_selection_grad} and
  Fig.\ref{trema_selection_grad}).} 
{ The main results are in line with the behaviours discussed
for the Poisson case: 
\begin{itemize}
\item for $\alpha=2$ the conditional density sensibly changes its
  shape corresponding to $\kappa(r) > 0.1$. For $\alpha=1$ the
  power-law behaviour of $\left< {n(r)} \right>$ is weakly affected
  { by selection effects}: the best fit exponent changes from
  $\gamma=-1.1$ to $\gamma=-1.3$. Instead, for $\alpha<0$ the
  difference with the unperturbed conditional density is manifested at
  large scales only.
\item   $\delta^p(r;h)$ is a very efficient diagnostic to
detect radial dependent effects as it shows a clear systematic behaviour as a
function of scale $r$ for different values of $\alpha$.  In particular when
$\delta^p(r;h)>0.1$ we observe a notable difference in the behaviour of the
conditional density in spheres, i.e. $\kappa(r) >0.1$.
\end{itemize}
}

It is interesting to note that $\delta^o(r;h)$ is now much more
fluctuating than { it was} for the Poisson case. These
fluctuations, which however are limited in amplitude, are caused by
both spatial structures {when $\alpha$ is small} and selection effects
when {$\alpha$ is large enough.}  Indeed, we recall that when the
cylinders are oriented orthogonally to the line of sigh passing for
their centres, at large $r$, they get contributions from points with
an increasing radial distance.

 { The radial density can be very easily changed} by a
  selection function of the type given in Eq.\ref{radselfunc}. 
  The use of the GCM, i.e. the analysis of $\delta^p(r;h)$
  and $\delta^o(r;h)$, is able to reveal the presence of a radial
  selection that is shown by a systematic trend in $\delta^p(r;h)$.

\subsection{Mock galaxy  catalogues} 

\begin{figure}[tbp]
\centering 
\includegraphics[width=1\textwidth]{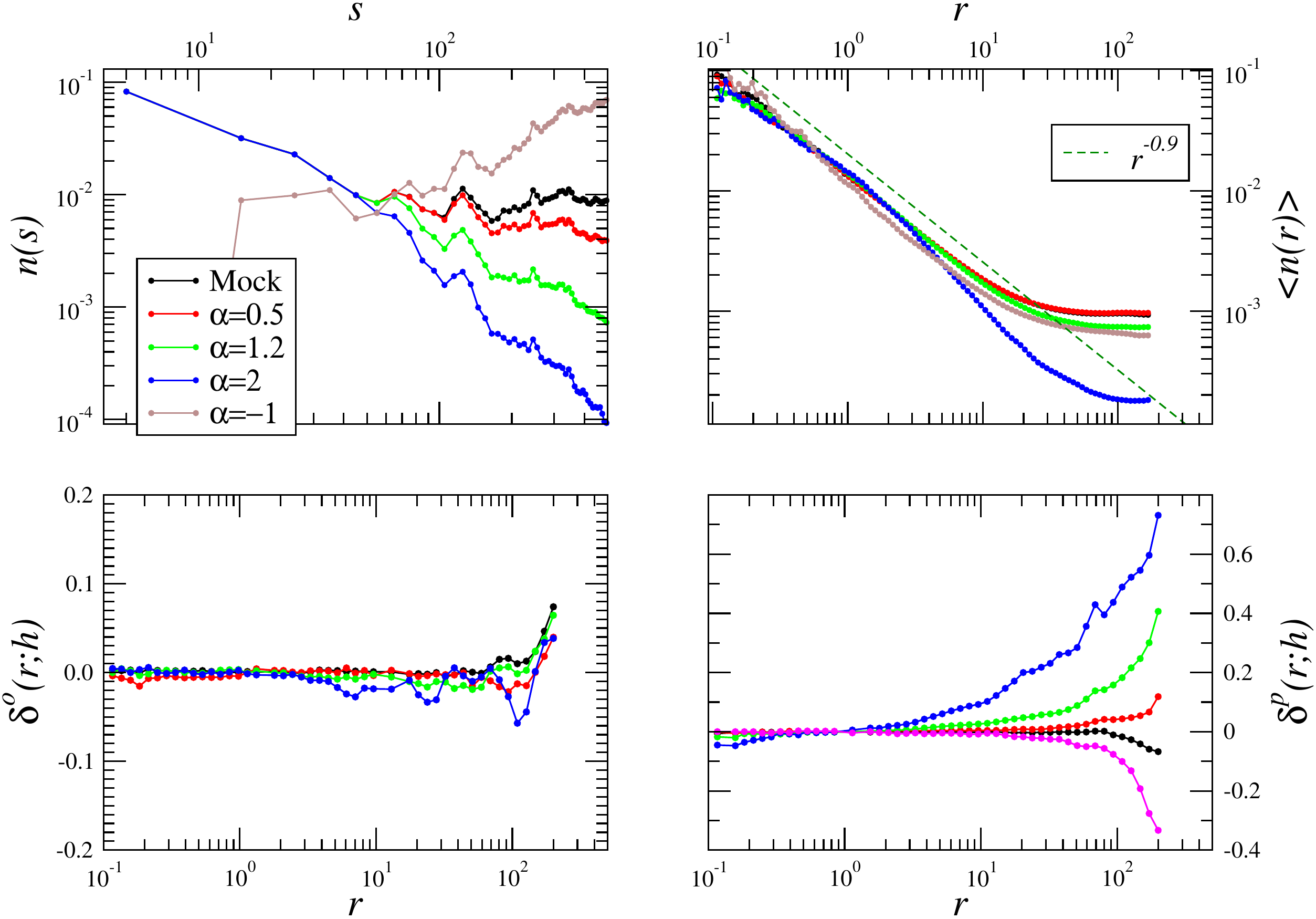}
\hfill
\caption{
As Fig.\ref{poisson_selection} but for the case of a mock
  galaxy catalogues.
\label{mock_selection_grad}
}
\end{figure}

A mock galaxy distribution, constructed by a cosmological LCDM N-body
simulation \cite{croton}, represents an intermediate situation
between a strongly clustered and a uniform distribution. Indeed, the
conditional density (see Fig.\ref{mock_selection_grad}) shows a
power-law behaviour at small scales, i.e. $r<20$ Mpc, while it flattens
at large ones. Correspondingly, at large enough distances, the radial
density presents fluctuations that are larger than a purely Poisson
case, but symmetric (on average) around a constant behaviour.  As for
the other cases previously considered, we note that the behaviour of
$\delta^p(r;h)$ is again a very good diagnostic of the presence of
radial selection effects: for $\alpha>0$ we have that $\delta^p(r;h)$
grows with the scale $r$ and for $\alpha<0$ we find that $\delta^p(r;h)$
decreases.

%%%%%%%%%%%%%%%%%%%%%%%%%%%%%%%%%%%%%%%%%%%%%%%%%%%%%%%%%%%%%%%%%%%%%%%%%%

\subsection{Tests on pair counting estimators} 

We have used the four estimators introduced above: (i)~the FS
(Eq.\ref{estimator_np}), (ii)~the integrated Rivolo (Eq.\ref{rivolo}),
(iii)~the DP (Eq.\ref{dp}) and (iv)~the LS (Eq.\ref{ls}) estimator to
compute the conditional density both in the case of the random walk
and of the trema dust. Note that for the three estimators based on
pair counting we used 10 times random points than in the original data
sample. { Galaxies and mock random samples have the same radial
  selection function.}

\begin{figure}[tbp]
\centering 
\includegraphics[width=.45\textwidth]{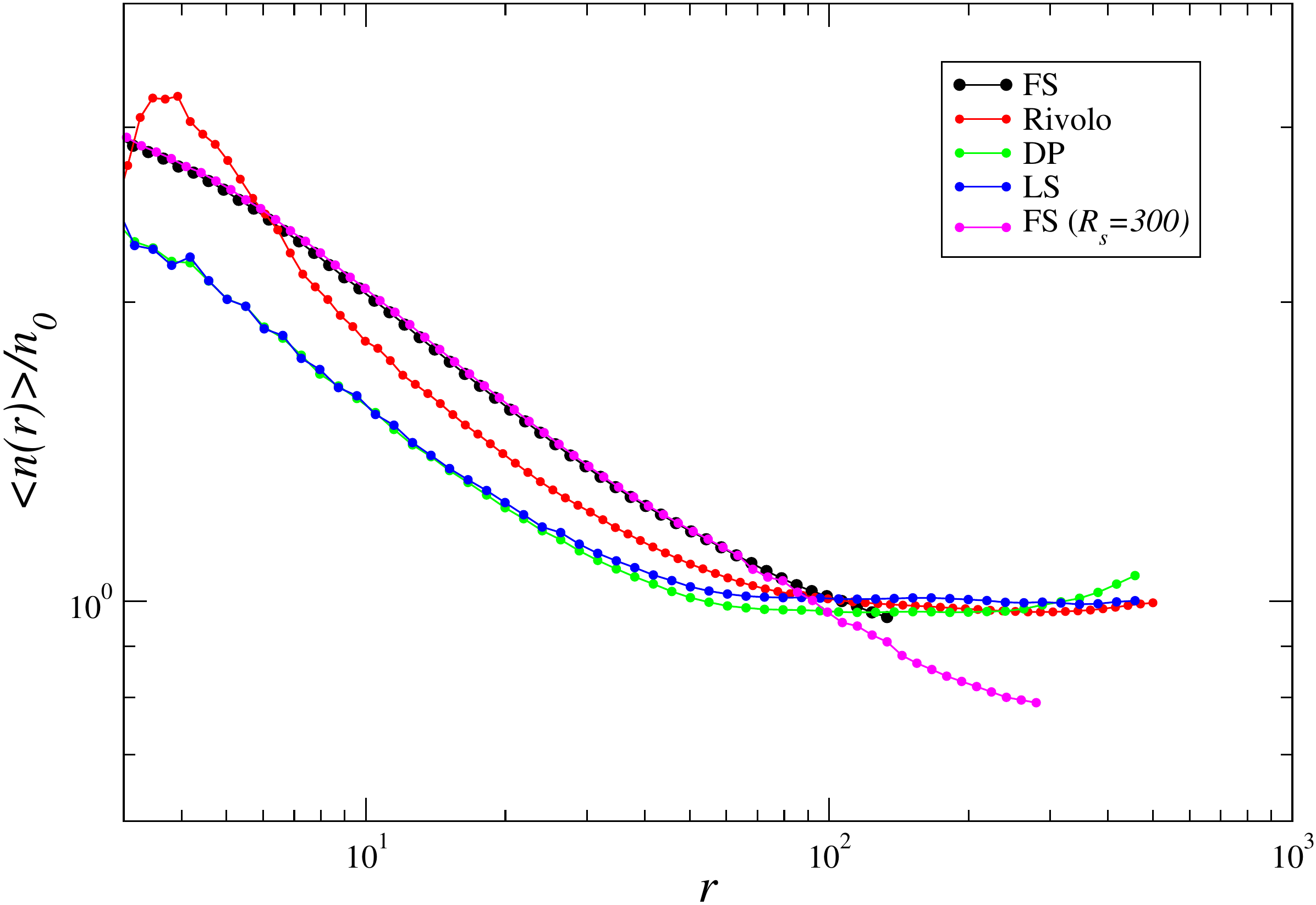}
\includegraphics[width=.45\textwidth]{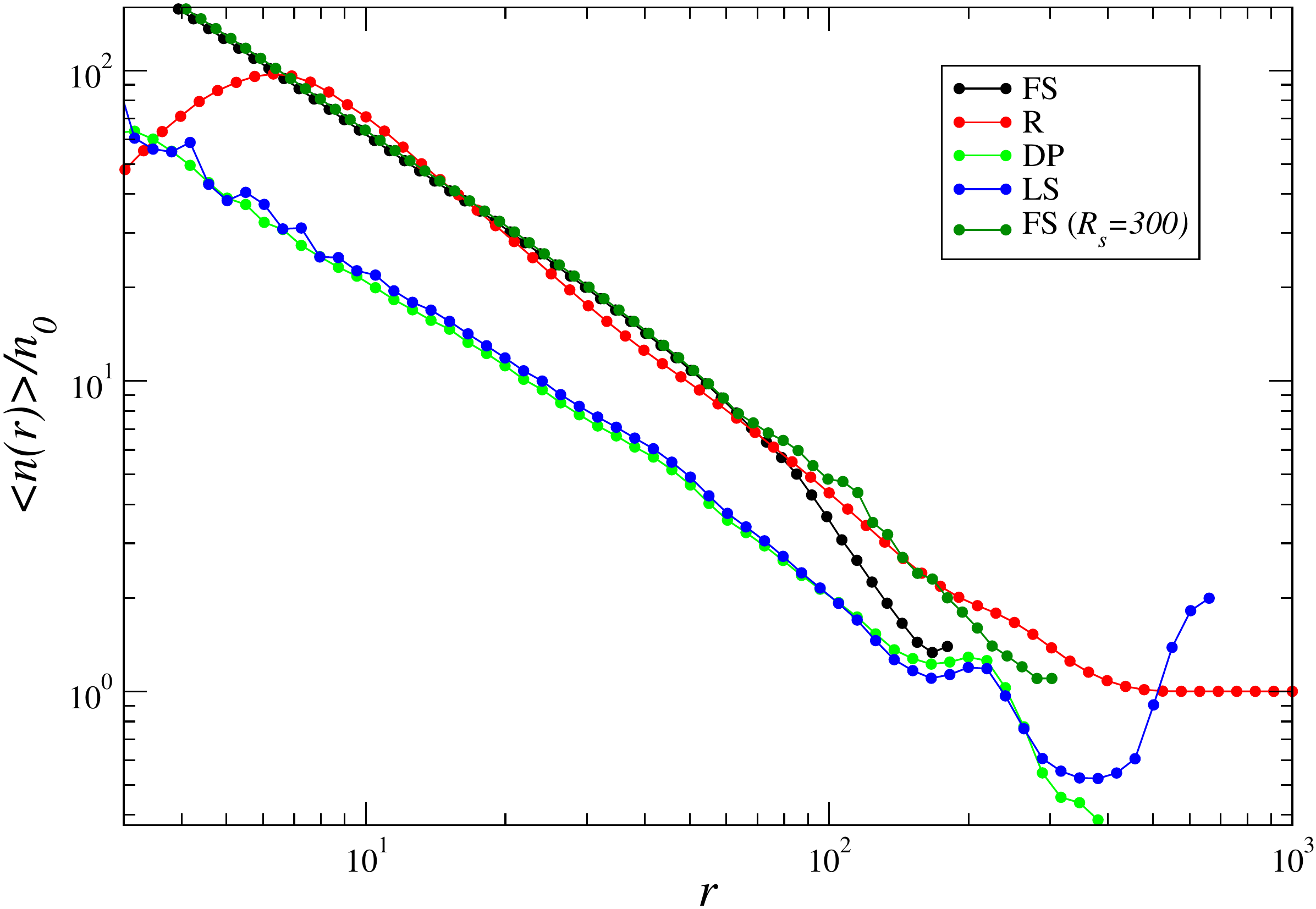}
\hfill
\caption{ Conditional average density computed for the trema dust with
  $D=2.7$ (left panel) and a random walk with $D=2$ (right panel)
  with different estimators: Full-Shell in spheres, Rivolo, Davis and
  Peebles,  Landy and Szalay. { Note that the estimator of the
    conditional density has been normalised to the value of the of the
    sample density $n_0=N/V$ (where $N$ is the total number of objects
    in the sample of volume $V$).}
\label{trema_spheres_paircounting}}
\end{figure}

 At small enough scales, i.e. $r \ll R_s \sim 150$ Mpc all estimators
 give similar results: { in particular, the slopes are similar.  The
   large finite size fluctuations that characterise a fractal
   distribution affect differently each estimator and thus the
   amplitude (once normalised to the sample average) is not the same
   in the different cases.  On scales $r \approx R_s$ or larger,
   i.e. the limit of the FS estimator, we find that the three
   pair-counting estimators show a clear flattening although the
   behaviour for $r>R_s$ is different for the two fractals with $D=2$
   and $D=2.7$ (see Fig.\ref{trema_spheres_paircounting}).}  {
   Therefore the large scale behaviour, i.e. $r \sim R_s$ and larger,
   of the conditional density is clearly completely spurious. } Even
 the FS estimator shows changes in its behaviour on scales comparable
 to $R_s$, where by construction there should be no changes in the
 underlying correlation properties:{ these are due to finite volume
   effects}.  { For comparison, we also show the FS estimator in a
   sample of double side $R^*_s = 2 R_S =300$: while the tail for $r
   \approx R^*_s$ is affected by finite size effects, at scales of the
   order of $R_s$ the FS estimator presents the expected behaviour}.

{  We note that other estimators of the correlation function than
  the FS one, based on pair-counting, are affected by systematic
  effects that are not analytically calculable. In particular, the
  different small amplitude is due to a different way of calculating
  the sample density: even the simplest estimator of the sample
  density 
\[
n_S = \frac{N}{V} \;,
\]
} { where $N$ is the number of galaxies contained in the sample
  volume $V$ is not an unbiased estimator, i.e. it does not satisfy to
  the condition that \be \langle n_S \rangle = n_0 \ee where $\langle
  ... \rangle$ is the ensemble average in the finite volume and $n_0$
  is the ensemble average \cite{book}. The bias in this measurement of
  the sample density depends thus on how the estimator is constructed
  and it enters as an {\it overall normalising factor}.  This thus is
  manifested even at small spatial scales. In addition, other
  systematic effects, which are different for different estimators,
  are manifested on scales of order of, or larger than, the radius of
  the maximum sphere that can be fully enclosed in the sample
  volume. In general, the smaller the fluctuations, and thus the
  larger is the fractal dimension, the smaller are the differences
  between the different estimators.  }

{The precise break down of the power law
   and the shape of the transition to ``uniformity'' depends on the
   convolution of the intrinsic correlation properties of the
   distribution with the window function of the sample.  In addition,
   different estimators show different large scale behaviours: the
   variance of each estimator cannot be controlled for arbitrary
   distributions and this is the reason why we use a conservative
   approach using the FS estimators (see also discussion in
   \cite{book}).}

We have performed other tests by considering mock LCDM galaxy samples
(see below), with or without a radial selection, instead of a highly
inhomogeneous one. We found that only in the case of LCDM point
distribution with a smooth radial selection of the type described by
Eq.\ref{radselfunc} one is able, by using a pair counting algorithm
and a random Poisson distribution with the same { radial distance
  counts} $n(s)$ of the data, to ``correct'' for the effect of the
selection function.

\begin{figure}[tbp]
\centering 
\includegraphics[width=.9\textwidth]{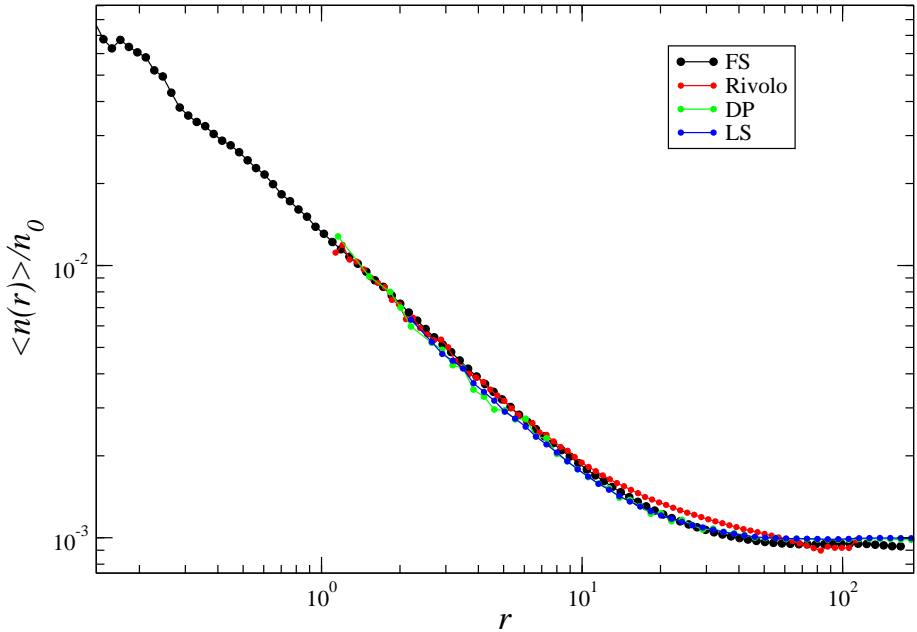}
\hfill
\caption{Conditional average density computed with different
estimators in the mock LCDM.}
\label{Lambda_paircounting}
\end{figure}

 {On the other hand, when the distribution is inhomogeneous any {
     measurement} of the conditional density (or of $\xi(r)$) with
   estimators that are not the FS gives a spurious results for
   $r>R_s$.}  { For this reason we conclude that estimations of the
   conditional density like those determined by \cite{Scrimgeour2012},
   cannot demonstrate that galaxy distribution is uniform but, at
   best, can reconstruct galaxy correlations if the distribution is
   uniform at scales much smaller than those of the sample.  This is
   the standard procedure adopted in three-dimensional clustering
   analyses, and it is based on the assumption that galaxy
   distribution is spatially uniform inside the given sample and that
   the cosmology close to LCDM.} 
{ Indeed, as shown in Fig.\ref{Lambda_paircounting}, the different
estimators of the conditional average density agree when they are
computed in a sample with small large scales fluctuations,
as it it the case of a realisation of a LCDM universe.}
However, as our aim is indeed to
   determine whether this is the case or not, we will not use an\
   other estimator than the FS one.

%%%%%%%%%%%%%%%%

\subsection{Discussion} 

To summarise { we have found that the conditional density is very
  weakly dependent on moderate selection effects in the data { both
    for spatially uniform and non-uniform point distributions}. Indeed
  only for a radial selection of the type $f(s) \sim s^{-\alpha}$ with
  $|\alpha| \ge 2$ the conditional density sensibly changes it shape.}
In addition, we have shown that the GCM is a very efficient tool to
reveal radial selection effects in data { working} equally well for a
uniform distribution and for an inhomogeneous one. { In particular
  the results is Sect.3 allow us to conclude that the measurement of
  the conditional density varies less than 10\%, i.e. $\kappa(r) < 0.1$
  (see Eq.\ref{Kappa}) when the gradient of the galaxy counts along
  the line of of sight (see Eq. \ref{deltap}) is $\delta^p <0.1$.}
{We stress that our method suggests that if a sample has
  $|\delta^p (r)| < 0.1$, and if the sample satisfy all other standard
  tests, like the obvious one of constructing a volume limited sample
  instead of considering a magnitude limited one, then the results can
  be trusted.}

{ Then, we have
  pointed out} that, in order to make a reliable estimation of large
two-point correlations one needs to use the FS estimator of the
conditional density. The main uncertainty of this estimator is
represented by systematic errors which are important at the scale of
the sample and of which we are not able to give a quantitative
estimation but only identify their presence through the study { e.g.,}
of the probability distribution of spatial fluctuations
\cite{sdss_aea}.

{ We note that the relation between the behaviours of $\delta^p(r;h)$
  and $n(s)$ is not uniquely determined. This is shown by the simple
  case of a Poisson distribution with a radial selection of the type
  $f(s) \sim s^{-1}$ and and a fractal with dimension $D=2$. In both
  cases we find the radial density decays as $n(s) \sim s^{-1}$, but
  only in the former case $\delta^p(r;h) \approx const.$. Thus, in
  general, it is not possible to predict the behaviour of the radial
  distance, or of the conditional density, from the knowledge of
  $\delta^p(r;h)$.  }
%

%%%%%%%%%%%%%%%%%%%%%%%%%%%%%%%%%%%%%%%%%%%%%%%%%%%%%%%%%%%%%%%%5

\section{Results on real galaxy samples}
\label{results} 

{ The samples for which we present the analysis in this section are
briefly discussed in Sect.\ref{data}. }

\subsection{Sloan Digital Sky Survey} 

We have considered three different sample of the Sloan Digital Sky
Survey (SDSS) \cite{york2000}: the Main Galaxy (MG) sample
\cite{strauss2002,paper_dr6}, the Luminous Red Galaxy (LRG) sample
\cite{lrg,kazin} and the Quasar (QSO) sample \cite{Schneider2010}.

\subsubsection{Sloan Digital Sky Survey Main Galaxy Sample}

We begin by discussing the behaviour of the FS estimator of the
conditional density, i.e. Eq.\ref{estimator_np}, in the six VL samples
of the Sloan Digital Sky Survey (SDSS) { (see Sect.\ref{mg-sample}
  and Tab.\ref{tab_sdss})}. In order to evaluate the value of the
cylinder radius, we have computed the nearest neighbours (NN)
distribution and { their average distance} $\Lambda$.  We found
that $\Lambda \approx 1-4$ Mpc/h for the various samples.  { Note
  that for MG VL samples we have tested the FS estimators of $\langle
  n(r) \rangle$ in cylinders converge to a stable estimation for $h
  \ge 1$ Mpc/h, while for the LRG samples (see below) this convergence
  occurs $h \ge 10$ Mpc/h}.

\begin{figure}[tbp]
\centering 
\includegraphics[width=1\textwidth]{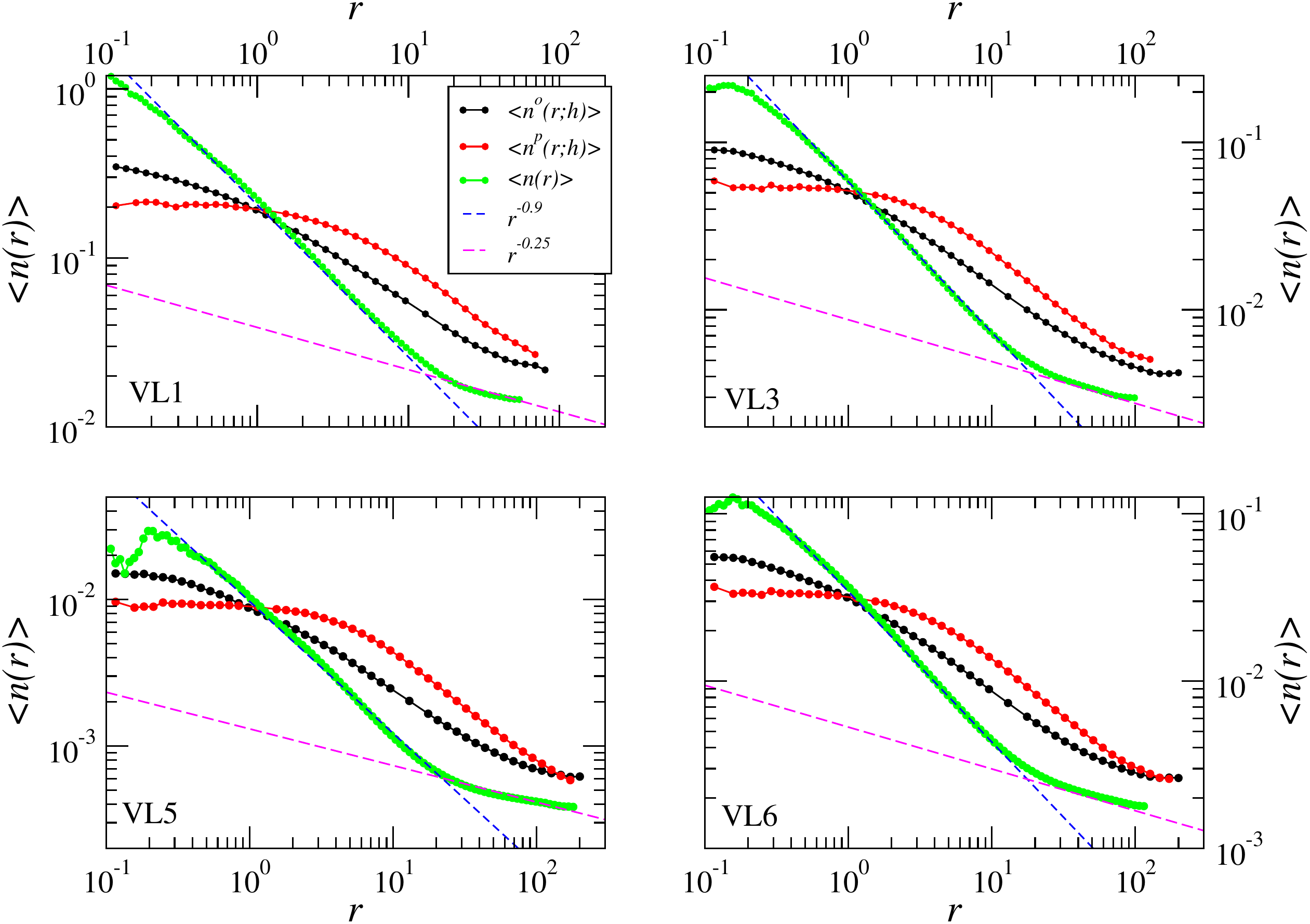}
\hfill
\caption{
Behaviour of the conditional density computed in spheres and
  in cylinders of radius $h=1$ Mpc/h in several VL samples of the SDSS
  survey.\label{sdss_cylinders}
}
\end{figure}

The behaviour of the conditional density is reported in
Fig.\ref{sdss_cylinders}.  One may note that in all samples we find a
consistent result for the FS estimator in spheres. { In particular we
  measure a power law behaviour $\sim r^{-\gamma}$ with $\gamma=0.9
  \pm 0.05$\footnote{ We performed a least-squares fit with equal
    weight on each point. The standard deviation refers to the fit in
    each sample.} in the range $[1,20]$ Mpc/h: the value of the slope
  and it error refer to the average over the six VL samples
  considered. At larger scales we find $\gamma=0.25\pm 0.05$ in the
  range $[20,110]$ Mpc/h .

Note that in this case the fits extends for less than a
  decade and thus one can easily find other possible fits with a
  functional form of the conditional density different from a simple
  power-law. However we conclude that at $\sim$ 100 Mpc/h a clear
  crossover toward homogeneity has not been yet reached.}
 As discussed in \cite{gumbel} the nontrivial scaling behaviour
of the average conditional density (and of its variance) correspond to
fluctuations obeying the Gumbel distribution of extreme value
statistics: we refer to \cite{sdss_aea,gumbel,cqg_review} for a more
extensive and detailed discussion of the variance and of the whole
probability distribution of fluctuations.

 {  The conditional density in cylinders $\left< n^p(r;h) \right>$
   and $\left< n^o(r;h) \right>$ show, for $r <10$ Mpc/h, show a
   different slope that can be interpreted as due to the effect of
   peculiar velocities. Indeed, because of the effect of peculiar
   velocities, structures, at small scales are more elongated along
   the line of sight than in any other direction. Thus in the range of
   scales where the deformation of peculiar velocities $v_p$ is
   relevant, i.e.  such that $r \approx v_P/H_0 \le 10$ Mpc/h, the
   conditional density along the line of sight is almost
   constant. Instead, on larger scales $\left< n^p(r;h) \right>$
   approaches, with a different amplitude, the same scale dependent
   behaviour as $\left< n^o(r;h) \right>$.}

{ Moreover, the conditional density in parallel cylinders, of
  $\left< n^o(r;h) \right>$, shows a change of slope at scales of the
  order of $\sim 100$ Mpc/h for $h=1$ Mpc/h. Whether this corresponds
  to a crossover toward homogeneity or a change of slope cannot be
  sorted out by this analysis. Indeed, as discussed in
  Sect.\ref{full-shell-estimator-in-cylinders}, when the conditional
  density in spheres is characterised by a change of slope (as the one
  described by Eq.\ref {twopower}) the large scale behaviour of $\left<
  n^o(r;h) \right>$ or of $\left< n^p(r;h) \right>$ does not clearly
  determine the correlation exponent as shown by the example in
  Fig.\ref{fig_1.0_0.0_20}.  }

\begin{figure}[tbp]
\centering 
\includegraphics[width=1\textwidth]{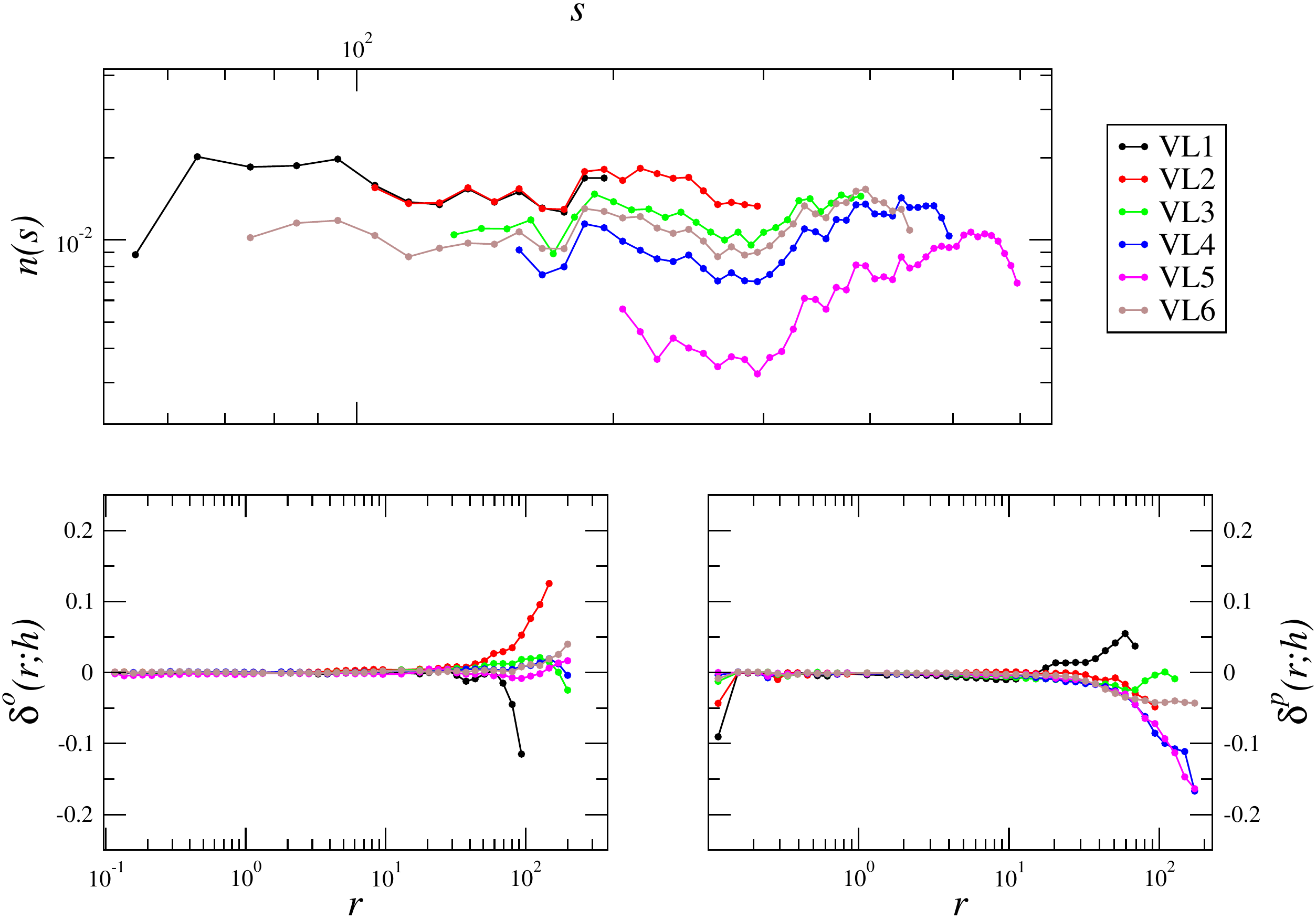}
\hfill
\caption{
As for Fig.\ref{sdss_ML_grad} but for the six VL samples of
  the SDSS survey: note that both the radial density $n(s)$ and the
  conditional density in spheres have been normalised by an { arbitrary
  factor}. 
\label{sdss_VL_grad}
}
\end{figure}

The behaviour of the radial density and of $\delta^o(r;h)$ and
$\delta^p(r;h)$ is reported in Fig.\ref{sdss_VL_grad}. One may note that
$\delta^o(r;h) \le 0.1$ but for the VL2 where, due to a large scale
fluctuation, it slightly increases its value. On the other hand
$\delta^p(r;h) \approx 0.1$ only for the VL4 and VL5: the decrease of
$\delta^p(r;h)$ for $r>100$ Mpc/h corresponds to the fact that the
radial density $n(s)$ increases for large radii. However this
increase does not seem sufficient to change the behaviour of $\langle
n(r) \rangle$ at large scales, whose behaviour is compatible with that
found in other samples.

The increase of $\delta^p(r;h)$ for $r>100$ Mpc/h corresponds to the
large differences found by \cite{sdss_aea,copernican} in the
probability distribution of fluctuations at large scales, i.e. $r>100$
Mpc/h, in these same samples. Whether or not a radial dependent
selection effect, as the significant evolution hypothesised by
\cite{loveday}\footnote{  Loveday  \cite{loveday} proposed to explain
  the apparent number density of bright galaxies increases by a factor
  $\approx$ 3 as redshift increases from $z= 0$ to $z= 0.3$ as with a
  significant evolution in the luminosity and/or number density of
  galaxies at low redshifts. However there are no independent proofs
  of this hypothesis than the observation of the growing number
  density observed in the SDSS samples\cite{sdss_aea}.}, contributes
to the behaviour of $\langle n(r) \rangle$ and $\delta^p(r;h)$ for
$r>100$ Mpc/h { in these samples cannot be definitively clarified from
  these data.}

{
Note that the effect of peculiar velocities on small scales, that is
shown by the difference in the behaviour between $\left< n^o(r;h)
\right>$ and $ \left< n^p(r;h) \right>$ on scales of order ten Mpc, is
not detectable in the analysis of $\delta^p(r;h)$ as on small scales
there are few points in the cylinders. One could increase the cylinder
radius $h$ but then taking $h=10$ Mpc/h one looses the small scale
effects as well.
} 

\begin{figure}[tbp]
\centering 
\includegraphics[width=1\textwidth]{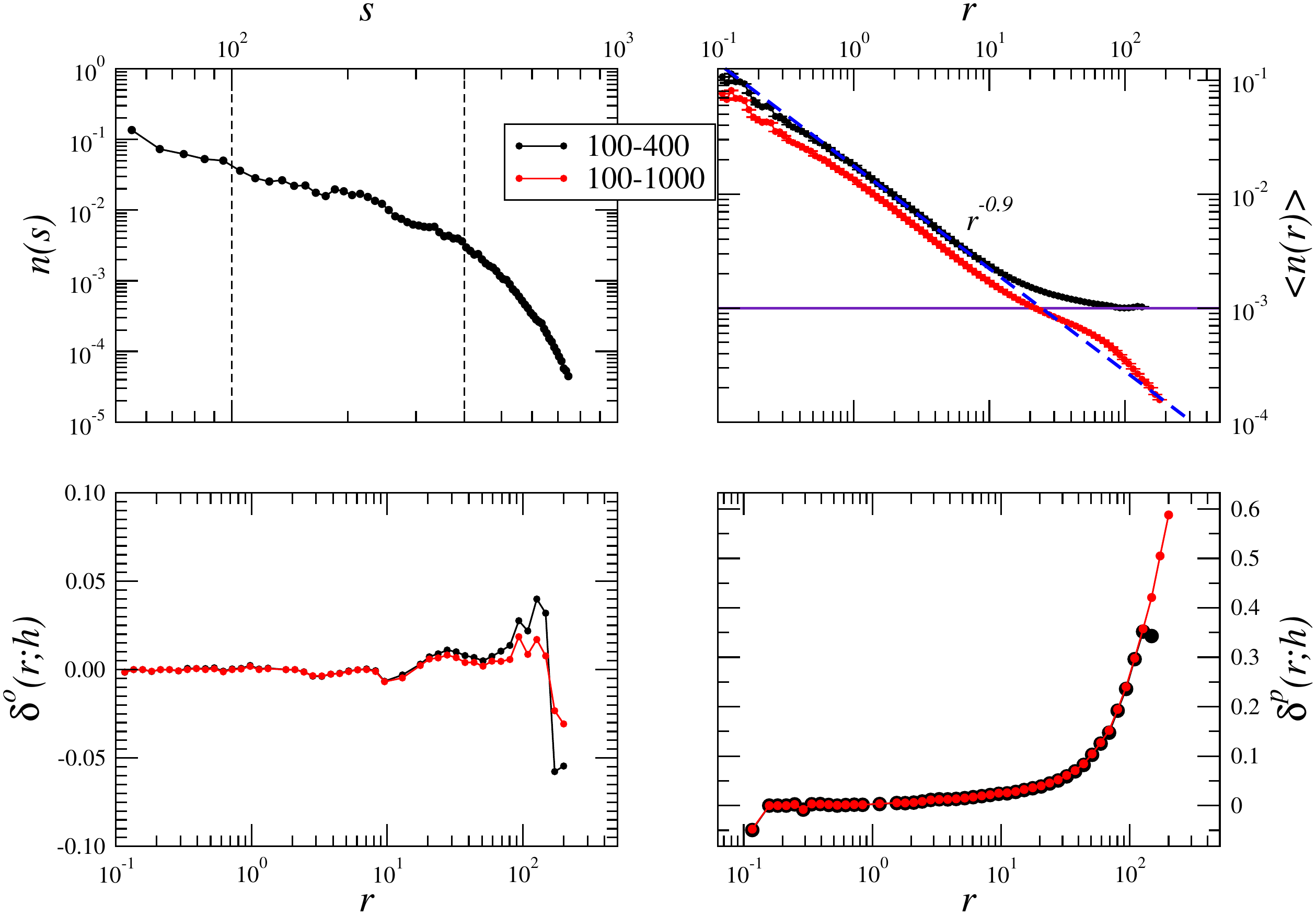}
\hfill
\caption{
The case of the SDSS-ML sample. Upper left panel: radial
  density $n(s)$. Upper right panel: conditional density $\left<
  {n(r)^{s}} \right>$. Bottom left panel $\delta^o(r;h=1)$
  (Eq.\ref{deltao}).  Bottom right panel: $\delta^p(r;h=1)$
  (Eq.\ref{deltap}).
\label{sdss_ML_grad}
}
\end{figure}

In order to show the effect of a large scale radial selection we have
studied the SDSS ML sample. In particular we have considered a
sub-sample limited by $r\in[100,400]$ Mpc/h and a sub-sample limited
by $r\in[100,1000]$ Mpc/h. Results are shown in
Fig.\ref{sdss_ML_grad}.  One may note that in the deeper sub-sample
$n(s)$ presents a sharp decay { related to the fact that at large
  enough distances only very bright galaxies are included in the
  sample. Correspondingly} the conditional density in spheres decays
{ faster than for the VL samples but this is spurious behaviour as
  shown by growth of $\delta^p(r;h)$ at large scales.} Instead, in the
former sub-sample $n(s)$ gently decay but $\delta^p(r;h)$ again shows
the effect of the biased luminosity selection. Note that, in this
case, $\langle n(r) \rangle$ flattens at $r>100$ Mpc/h: this is a
spurious effect due to the (known) radial selection affecting the ML
sample. At least, in the VL samples that are not affected by the
stronger luminosity selection effect, we do not detect such a clear
flattening.

\subsubsection{Sloan Digital Sky Survey Luminous Red Galaxies Sample}

\begin{figure}[tbp]
\centering 
\includegraphics[width=1\textwidth]{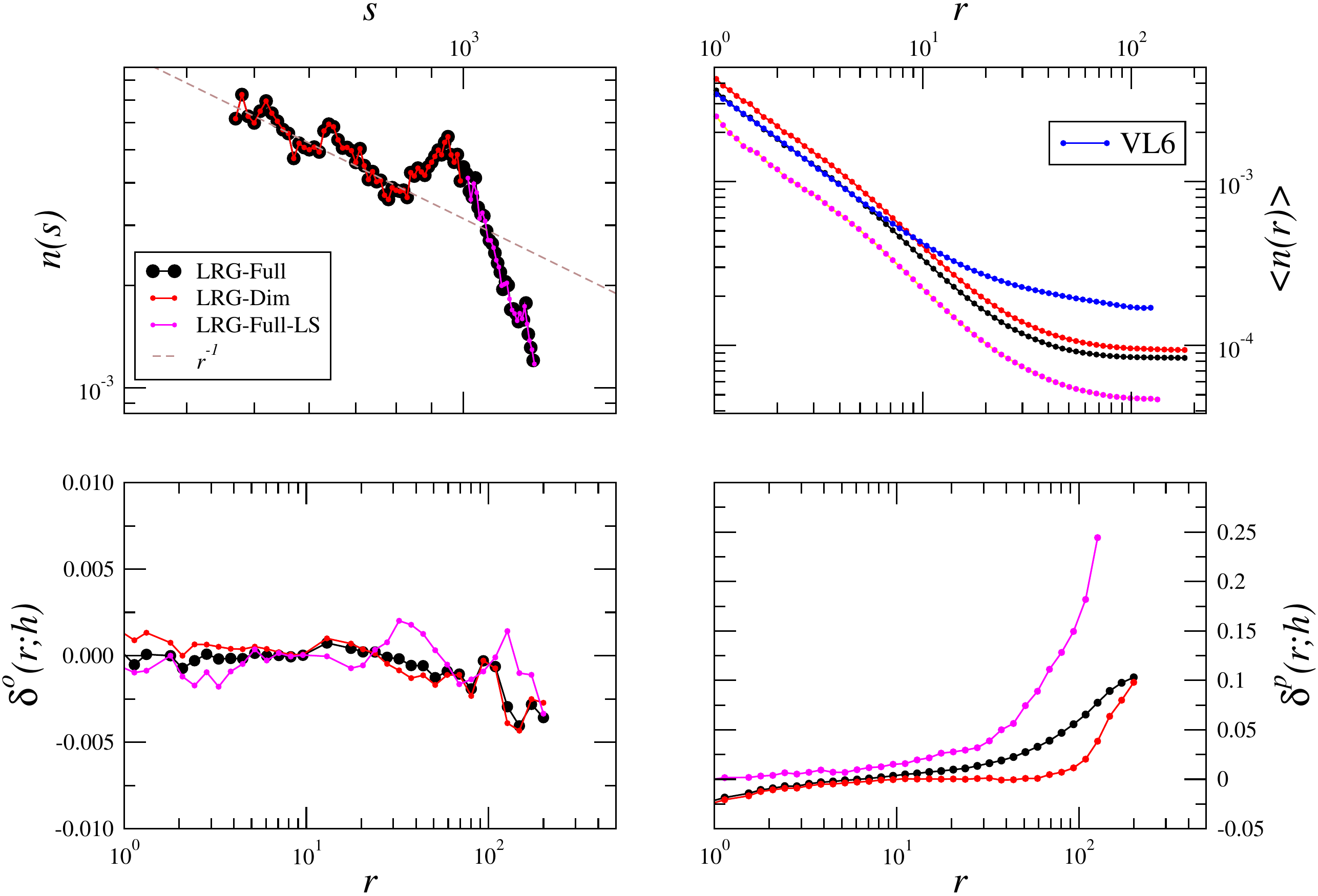}
\hfill
\caption{
As for Fig.\ref{sdss_ML_grad} but for the three LRG samples
  of the SDSS survey (the cylinder radius is $h=10$ Mpc/h). For
  comparison we reported the FS estimator of the conditional density
  in sphere for VL6-SDSS (see Fig.\ref{sdss_cylinders}).
\label{sdss_LRG_grad}
}
\end{figure}

In Fig.\ref{sdss_LRG_grad} we show results for the LRG samples. In
particular, we considered the Dim and Full samples and, for
comparison, the Full sample limited by radial distances in the range
$[10^3,1.5\times 10^3]$ Mpc/h --- we call this sample Full-LS {
  (see Sect.\ref{lrg-sample})}. This latter was chosen to see clearly
the effect of the large scale selection effect present in these
data. One may note the radial density decays as $\sim r^{-1}$ up to
$\approx 800$ Mpc/h, lowering its amplitude by a factor $\approx 3$,
then it grows by a factor $2$ in the range $800-1000$ Mpc/h and
{finally} for $r>1000$ Mpc/h it decays sharply. As mentioned above
this latter decay is due a {known} luminosity selection effect. The
question is whether the other large scale trends are due to other
selection effects, and thus to the fact that the sample is quasi VL,
or to intrinsic fluctuations.

{  %
The conditional density in spheres for the LRG samples approaches
  an almost constant value for $r>80$ Mpc/h.
However $\delta_p$ systematically increases at scales larger than
$\sim 50 \div 100$ Mpc/h with an amplitude that depends on the
different magnitude/redshift cut used.
Clearly in the LRG-Full-LS we observe the more clear trend with the
largest value of $\delta_p \approx 0.1$ at 100 Mpc/h. This is
certainly related to the sharp break of $n(r)$ beyond 1000 Mpc/h. On
the other hand the LRG-Dim sample shows $\delta_p \approx 0.1$ only at
$\approx 100 Mpc/h$ and the same occurs to the LRG-Full which is the
sum of the previous two samples. While the growth for LRG-Full for
scales $r>100$ Mpc/h is again related to the sharp break of the radial
density for $r>1000$ Mpc/h, present also in this sample, it is unclear
why the LRG-Dim sample shows a systematic increase in the range
[100,200] Mpc/h as in this case there are not known selection effects
at work. Therefore given that we are unable to draw a clear
identification of selection effects in this sample and because beyond
$\sim 10$ Mpc/h the behaviour of $\langle n(r) \rangle$ is different
in the LRG samples and in the sample MG-VL6 our conclusion is that we
need a larger sample to confirm the the transition of homogeneity at
$80$ Mpc/h that was claimed to be identified by \cite{Hogg2005}.  It
is worth to stress that our measurements, but not our interpretation
for the reasons discussed above, agree with the results of
\cite{Hogg2005}. }

\subsubsection{Sloan Digital Sky Survey Quasar Sample}

\begin{figure}[tbp]
\centering 
\includegraphics[width=1\textwidth]{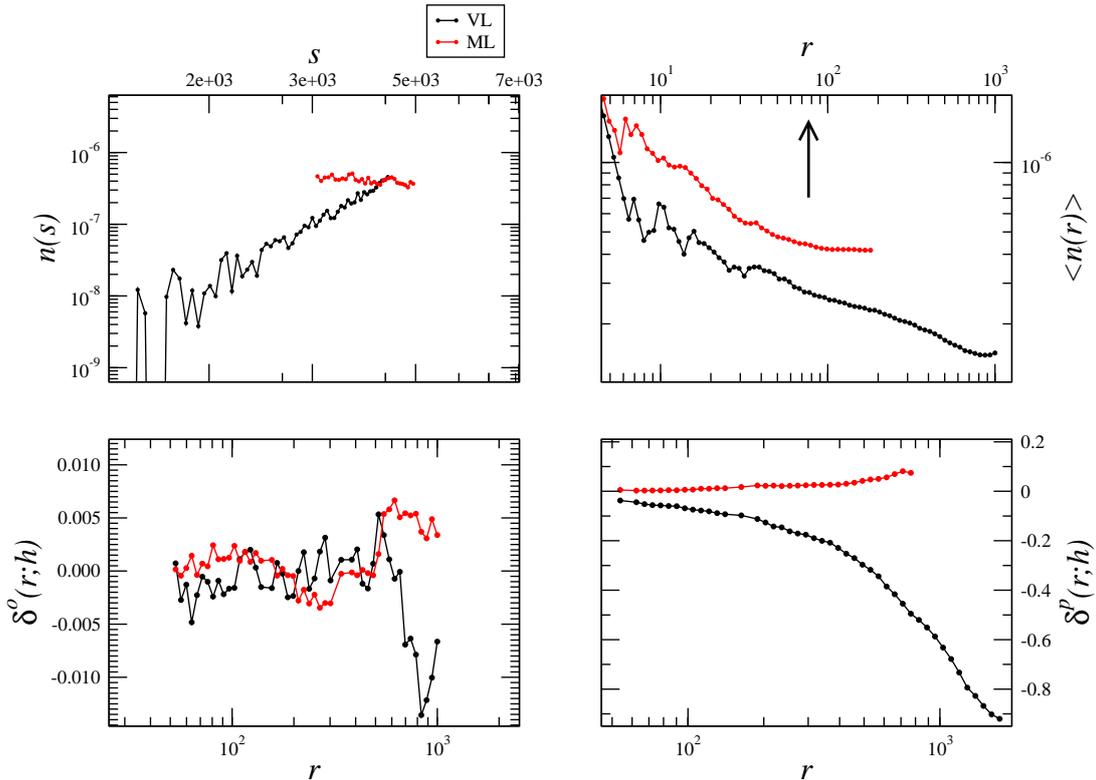}
\hfill
\caption{
As for Fig.\ref{sdss_ML_grad} but for the VL and the ML SDSS
  quasars samples. In this case and with $h=75$ Mpc/h. The arrow
  indicates the average distance between nearest neighbours.
\label{sdss_QSO_grad}
}
\end{figure}

 {The behaviours of the various statistics for the QSO SDSS samples {
     (see Sect.\ref{qso-sample} for more details)} are shown in
   Fig.\ref{sdss_QSO_grad}}. One may note that the VL sample is
 characterised by a major radial selection effect which dominates the
 estimation of correlation properties. On the other hand, { in the ML
   sample used by \cite{clowes2013,nadathur2013} the radial density is
   approximately constant: this behaviour cannot be, however,
   interpreted as an evidence in favour of spatial homogeneity for the
   following reasons.}

{ For a static population of uniformly distributed objects one
  expects to find: (i) a constant density in a VL sample and (ii) a
  density distributing $n(z)$ which reflects the selection function of
  the survey, i.e., that decays on large enough scales as for the
  magnitude limited sub-sample of the MG sample discussed above. In the
  VL-QSO sample we find instead that $n(z)$ increases at higher
  redshifts while it is almost constant in the QSO-ML sample. These
  peculiar behaviours occur because there must a strong selection
  effect intervening in these samples. This is related to the fact
  that properties of quasars are known to evolve with time because of
  the rate of mergers etc., and the redshift interval is very large,
  corresponding to more than 2 Gyr difference in look-back time. For
  this reason neither the VL nor the ML sample provide a sample of
  similar objects.}

{  The QSO ML sample has thus a nearly constant density as a
  function of redshift: this is a coincidence probably due to QSO
  evolution. We thus conclude that this QSO sample is still not
  suitable for measuring QSO correlation properties in three
  dimensions. On the other hand studies (see, e.g.,
  \cite{Shen_2008,Ross_2009}) using QSO samples for the derivation of
  correlation properties must necessarily adopt a number of ad-hoc
  assumptions to treat the redshift-dependent trends that characterise
  these samples. Here we have shown that these assumptions play a
  central role for deriving any result.}

  {Note that the GCM
  we have introduced is tuned to detect simple radial dependent
  selection effects. However for the QSO samples the GCM alone is not
  able to clarify the situation as this is more complex than for a low
  redshift galaxy samples. Indeed, in this case, in addition to a
  luminosity selection effect which is by construction present in the
  magnitude limited sample, the QSO are affected by a redshift
  dependent physical effect, most probably evolution. Thus in order to
  understand what is going on in these data, one has to combine the
  GCM with the analysis of the radial density in magnitude and volume
  limited samples. The results of this analysis is that, although
  $|\delta^p (r)| < 0.1$, these sample are affected by large redshift
  dependent selection effects, that are both observational (e.g., the
  magnitude limited selection) and physical (QSO evolution).}

{
The case of the QSO-ML sample shows indeed that our new criteria is
  complementary to the standard way to define a proper sample for
  statistical analysis. Indeed in addition to have $|\delta^p (r)| <
  0.1$, one should always include the well-known and obvious procedure
  of avoiding to use a magnitude limited sample. }

\begin{figure}[tbp]
\centering 
\includegraphics[width=1\textwidth]{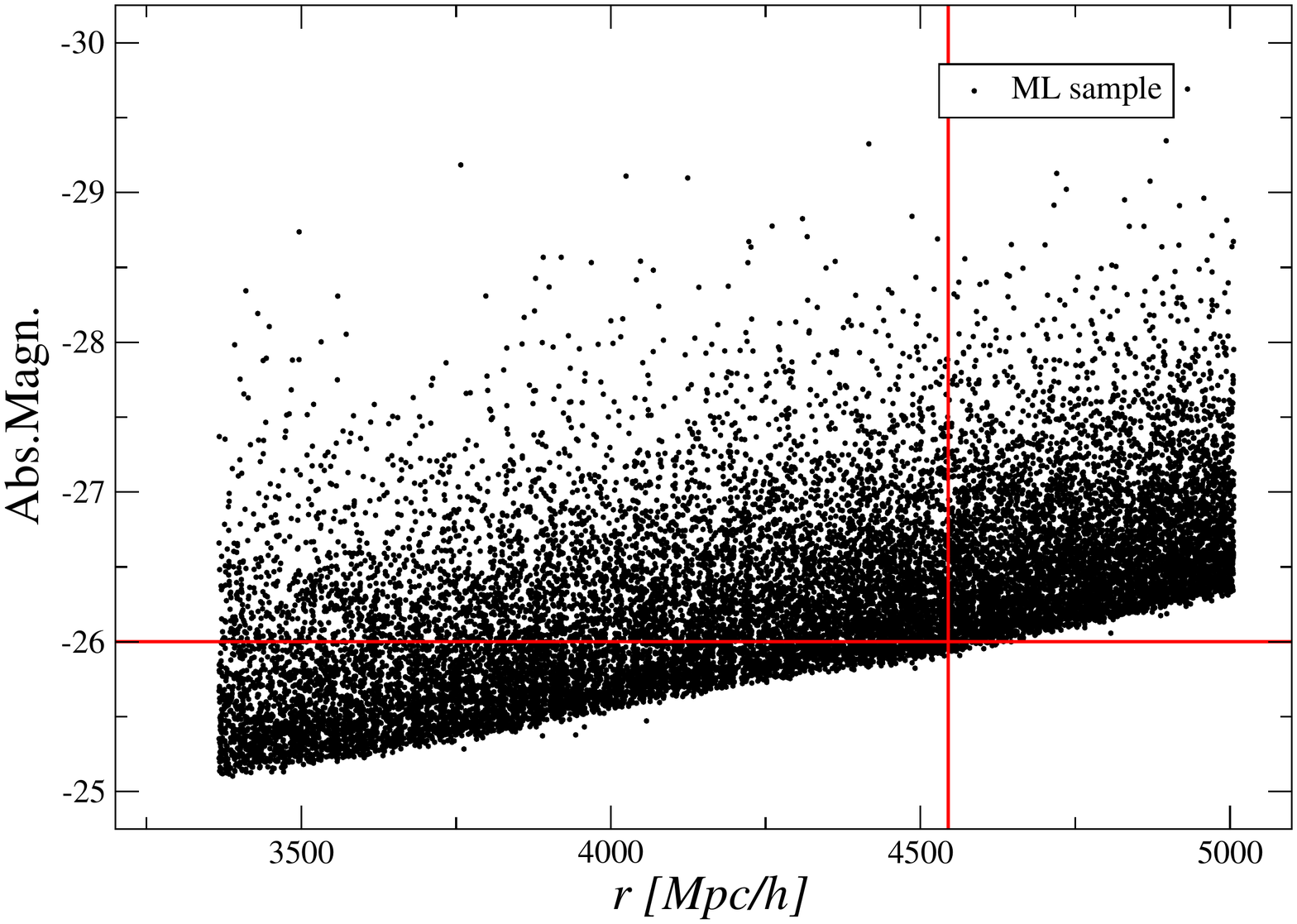}
\hfill
\caption{ Distance-absolute magnitude diagram for the Sloan Digital
  Sky Survey Quasar Sample. One may note that the full magnitude
  limited sample is affected by a major selection effect that
  introduces a bias in the luminosity of objects as a function of
  their distance. That is, intrinsically bright object are visible at
  all scales while intrinsically fain object are included in the
  survey only if their distance is small enough. Such a situation
  makes unsuitable the magnitude limited sample for a correlation
  analysis. In the diagram a VL sample covers a rectangular area. 
\label{sdss_QSO}
}
\end{figure}

%\begin{figure}[tbp]
%\centering 
%\includegraphics[width=1\textwidth]{fig13c.pdf}
%\hfill
%\caption{ Nearest neighbour distribution for the ML and VL QSO
%  sample.  The mean distance is at about $\sim 75$ Mpc/h. 
%\label{sdss_QSO_NN}
%}
%\end{figure}

\subsection{The Two Degree Field  Galaxy Redshift Survey}

{ In this and in the following section we have consider two galaxy
  redshift surveys which cover a smaller volume than SDSS but they
  allow an independent determination of the conditional density,
  although at a smaller scale than SDSS, in different sky regions and
  for different galaxies, as the selection properties of these surveys
  are very different from SDSS. }

Results for the Two Degree Field Galaxy Redshift Survey (2dFGRS) VL
samples {(see Sect.\ref{2dfgrs})} are shown in Fig.\ref{2df_VL_grad}.
One may note that the radial counts $n(s)$ show a sequence of
fluctuations, that correspond to large scale structures and voids.
Indeed, $\delta^p(r;h)$ does not present any large scales trend, while
$\delta^o(r;h)$ fluctuates at very large scales because of the {
  intrinsic fluctuations (structures) present in the sample NGC400.}
The conditional density in spheres present a $\sim r^{-0.9}$ decay
compatible with the results in SDSS, although the change of slope at
$\approx 20$ Mpc/h is less evident: {this is probably due to the
  weaker statistics, { i.e. the 2dFGRS samples cover a smaller
    volume and contain a lower number of points than the SDSS
    samples.}

\begin{figure}[tbp]
\centering 
\includegraphics[width=1\textwidth]{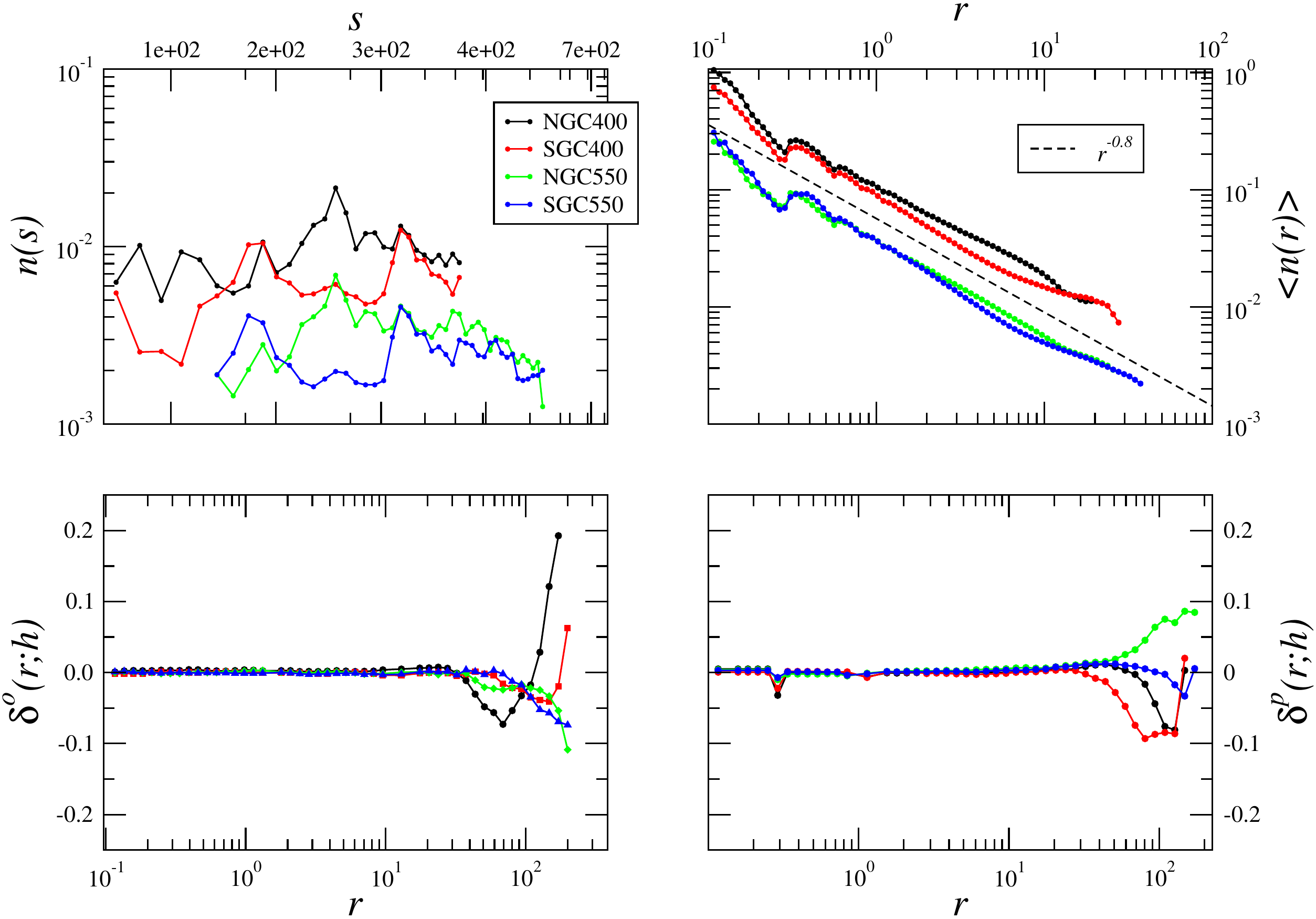}
\hfill
\caption{
As for Fig.\ref{sdss_ML_grad} but for the 4 VL samples of
the 2dFGRS survey ($h=1$ Mpc/h).
\label{2df_VL_grad}
}
\end{figure}

The behaviours of the FS estimators of the conditional density in
spheres and cylinders are reported in Fig.\ref{2df_cylinders}.  In
this case we observe that there is almost a factor ten between the
distance up to which we may compute the conditional density in spheres
and in cylinders. Such a difference is due to the geometry of the
sample, i.e. 2dFGRS cover a small solid angle in the sky but it is
relatively deep. However the large scale behaviours of $\langle n^p(r;h)
\rangle$ and of $\langle n^o(r;h) \rangle$ do not allow to make a
reliable estimate of the large scales, i.e. $r> 50$ Mpc/h, correlation
exponent:  from the one hand there are significant
  sample-to-sample fluctuations and from the other hand, as discussed
  above, if there occurs a change of slope at $\lambda_*\approx 20-40$
  Mpc/h it is not possible to make a reliable estimation of the
  exponent with in the conditional density { in cylinders.}

\begin{figure}[tbp]
\centering 
\includegraphics[width=1\textwidth]{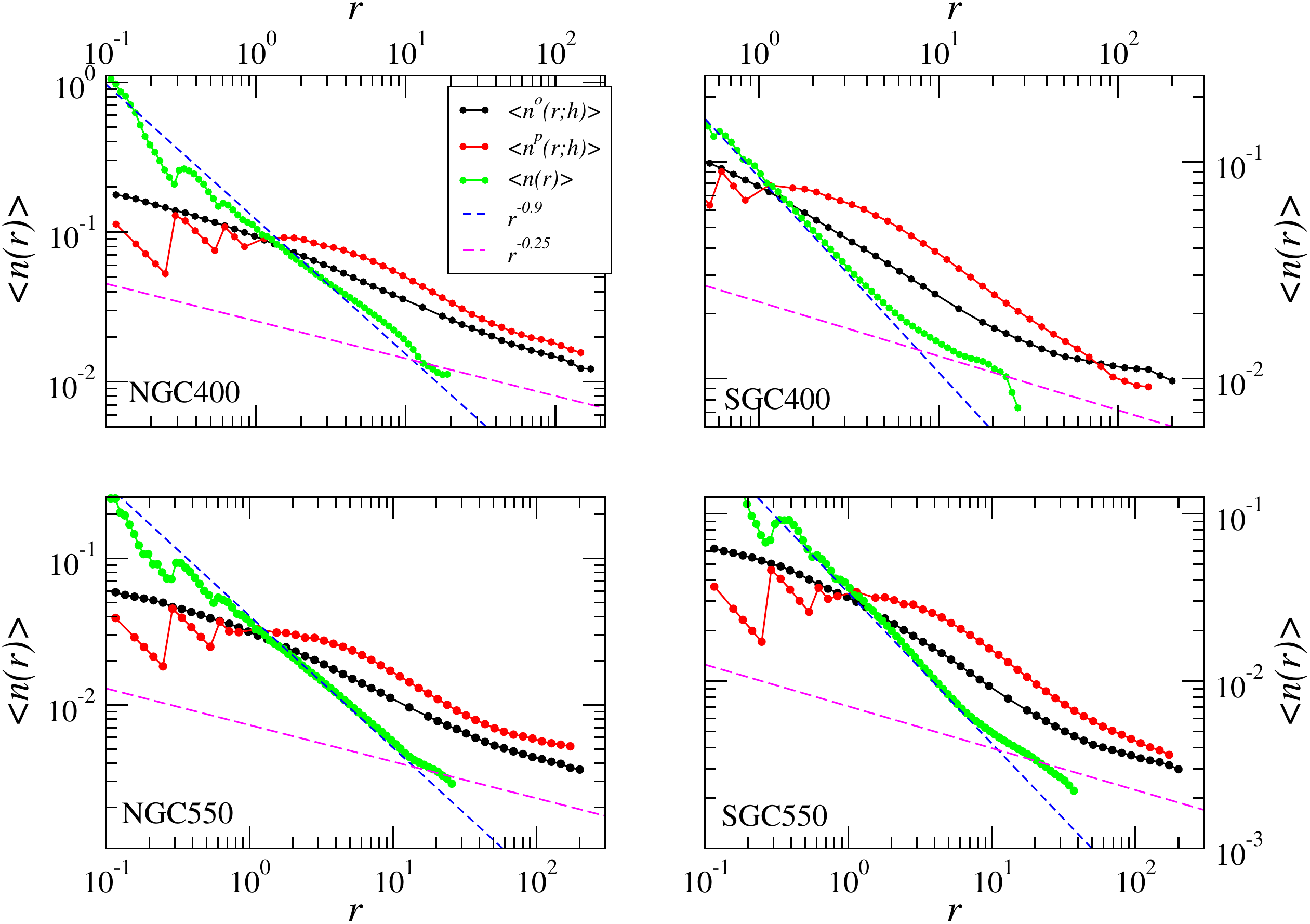}
\hfill
\caption{Behaviour of the conditional density computed in spheres and
  in cylinders of radius $h=1$ Mpc/h in several VL samples of the 2dFGRS
  survey.
\label{2df_cylinders}
}
\end{figure}
%

%%%%%%%%%%%%%%%%%%%%%%%%%%%%%%%%%%%%%%%%%%%%%%%%%%%%%%%%%%%%%%%%

\subsection{The Two Micron All Sky Galaxy Redshift Survey}

The various samples of the Two Micron All Sky Galaxy Redshift Survey
(2MRS) samples { (see Sect.\ref{2mass})}, { that contains galaxy
  selected in the near infrared at low redshift and thus with small
  evolutionary and K-corrections}, do not show any { strong} radial
selection effect (see Fig.\ref{2mrs_grad}). The conditional density,
at small scales, shows a power-law behaviour that is similar to the one
detected in the VL samples of SDSS and 2dFGRS. Even in this case it is
not possible to investigate further the change of slope at large
scales because of the limited size of the samples and the relatively
small number of points contained.

\begin{figure}[tbp]
\centering 
\includegraphics[width=1\textwidth]{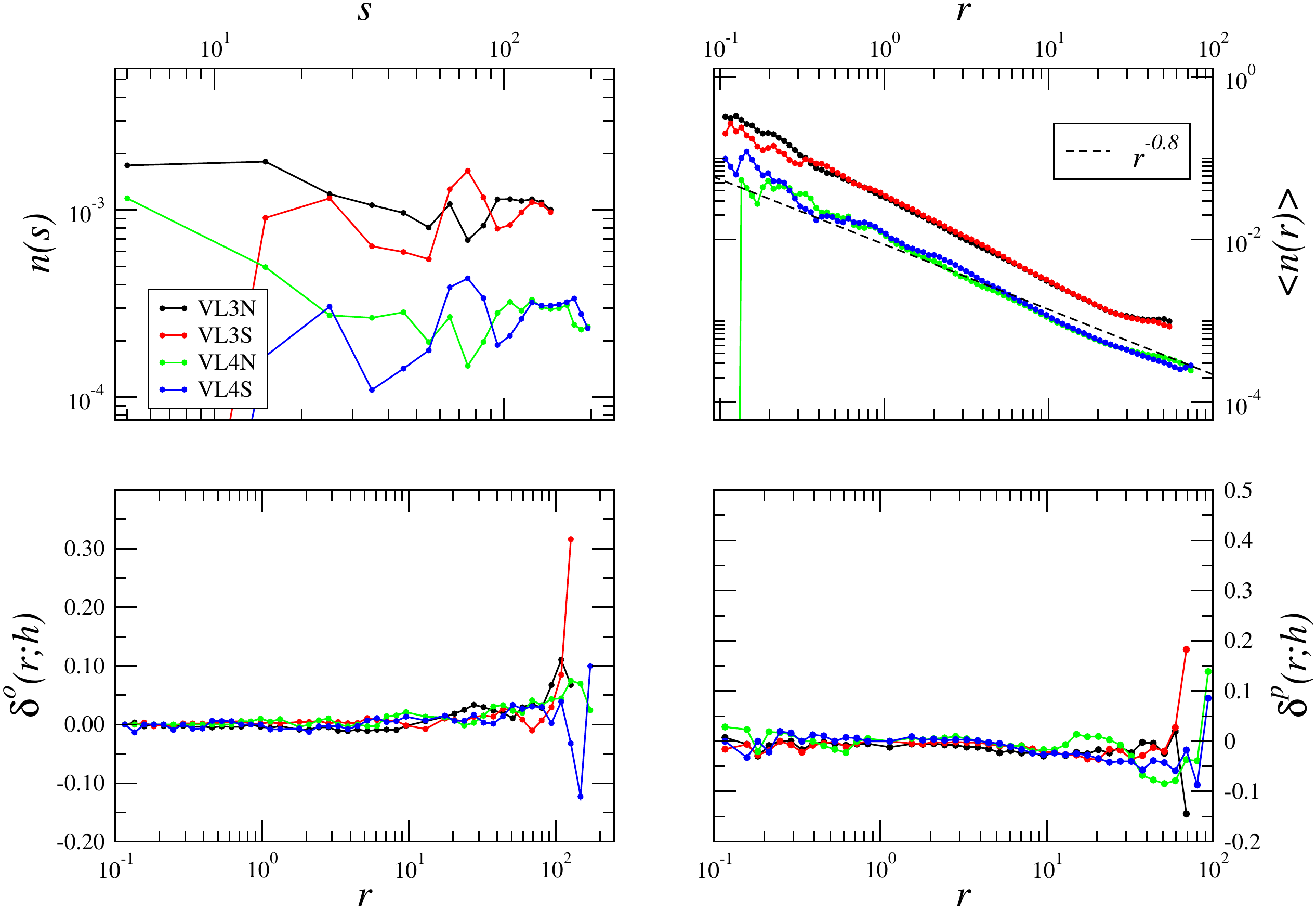}
\hfill
\caption{ As for Fig.\ref{sdss_ML_grad} but for the 4 VL samples of
  the 2MRS survey ($h=1$ Mpc/h).
\label{2mrs_grad}
}
\end{figure}

\section{Discussion and Conclusions} 
\label{discussion}

The simplest statistical quantity to characterise spatial correlations
of a point distribution is represented by the conditional density
$\langle n(r) \rangle$. This gives the average number of points {
  in a sphere of radius $r$ (or in a shell of thickness $\Delta r$ and
  radius $r$)} centred in a distribution point \cite{book}. The
standard two point correlation function is related to the conditional
density but it requires the additional estimation of the sample
density $n_0$, as $\xi(r)= \langle n(r) \rangle/n_0 -1$. This
statistics gives a meaningful physical result only if the estimation
of sample average represents a reliable estimation of the ensemble
average density, i.e. only if the distribution is uniform well inside
the sample volume. Therefore, in order to test whether a distribution
is indeed uniform inside a given sample one should therefore study the
conditional density and only if this show a scale independent behaviour
one can then consider $\xi(r)$ to characterise correlation properties
of small amplitude density fluctuations.

 {Given this situation the next problem we considered was to clarify
which is the most suitable estimator of $\langle n(r) \rangle$. There
are several studies discussing the problems related to the different
estimators introduced in the literature (see, e.g.,
\cite{kerscher1999,kerscher00,book,cdm_theo}).}
They all compute the number of points in spherical volumes and the
basic distinction is whether or not these volumes are fully included
in the sample. The full shell (FS) estimator { uses} only complete
spherical volumes, while non-FS estimators use volumes which are not
completely included in the sample boundaries, by giving appropriate
weights to them.  It was argued in various previous works { (see,
  e.g., \cite{kerscher1999,book} and references therein) } that the FS
makes the most conservative estimation, limiting the analysis to the
radius $R_s $ of the largest sphere fully contained in the sample
volume: { for deep surveys, when the survey solid angle is small,
  this can that can be much smaller than the maximum distance
  $R_{max}$ between two points.  However, the advantage is that, when
  the statistical volume average is properly performed, i.e. for
  scales smaller than $R_s$, there are not unknown biases affecting
  the estimation.}

 { Instead non-FS estimators, like the \cite{dp83} the \cite{ls} and
   the \cite{rivolo} estimators, can reach scales of the order of
   $R_{max}$. However, for a generic distribution, they are affected
   on large scales, i.e. $R_s \le r \le R_{max}$ by strong finite size
   effects \cite{kerscher1999,book}}. We have shown, { by
   considering the determination of non-FS estimators for several
   point distributions with a-priori know properties (see
   Sect.\ref{toys}), that this is indeed the case}.  For this reason
 we conclude that the claim of a transition toward uniformity by
 \cite{Scrimgeour2012} is valid only under the assumption that the
 distribution is uniform inside the WiggleZ survey.  Indeed, the
 WiggleZ having a complex angular and redshift selection function,
 does not represent a suitable sample for an analysis with a FS
 estimator { which instead requires a contiguous volume
   corresponding to a uniform sampled sky area.}
{ In order to take into account these problems, the Rivolo
  \cite{rivolo} estimator was used by authors of the WiggleZ survey
  correlation analysis \cite{Scrimgeour2012}. However, this estimator
  is affected, for a generic distribution,} by an unknown bias and
variance on large scales. { For this reason the result obtained by
  \cite{Scrimgeour2012} } represent only a self-consistency test of
the transition to homogeneity rather than a robust model-independent
test of homogeneity itself. That is, if galaxy distribution is uniform
at small scales., i.e. $r\approx 50$ Mpc/h, then with the Rivolo
estimator is able to improve the estimation of the conditional
density (or of $\xi(r)$) by taking into account the complicated
selection function of the survey.  However if galaxy distribution is
not uniform by using the Rivolo estimator one cannot draw any
definitive conclusion on the value of the large scale correlation
exponent.

{ We have discussed in detail above that} the FS estimator has
  another important limit: it can be applied only to samples which do
  not suffer for relevant radial and angular selection effects. Thus
  first of all one must use volume limited (VL) samples, that are not
  affected by the Malmquist bias \cite{BarTee12}, and then one should
  control that no other radial {and/or angular} selection effects are
  present in the samples. { For what concerns the latter, the
    conservative procedure, that we have used in this paper, to take
    into account the inhomogeneous angular sampling is to limit the
    analysis only to that part of the sample where the angular
    completeness is $>90\%$ and as uniform as possible.} This is not
  always possible and for instance catalogues like WiggleZ
  \cite{Drinkwater2010} or the first data release of the BOSS sample
  (i.e., SDSS DR9) \cite{boss} have a very inhomogeneous angular
  coverage of the sky, a fact that makes them unsuitable for a
  correlation analysis with the FS estimator.

Instead, the non FS estimators are able to correct selection effects
by comparing data-data counts to the data-random points counts, where
the random catalogues are characterised by the same radial { and
  angular} selection as the real data. This means that these random
catalogues have the same radial counts $n(s)$ as the real data. However,
this correction method is valid only for distributions which are
uniform well inside the sample boundaries. In this way, one implicitly
assumes that any departure of $n(s)$ from an almost constant behaviour
on large scale is due a selection effect.

The conclusion of this discussion is that the non FS estimators should
not be used to test whether a distribution is uniform inside a given
as, for different reasons, they work under the assumption that a
distribution is indeed uniform inside the given sample.  There are
then two open problems: (i) how to extend the analysis beyond $R_s$
and whether this is at all possible and (ii) how to identify and
possibly quantify spurious radial selection effects in VL samples
introduced by observational and/or  physical reasons.

In order to reach separations of the order of $R_{max}$, { i.e., of
  the order of} the maximum distance between two points in the survey,
with { the} FS estimator we have measured the conditional density { in
  cylinders}, included in the sample volume. The cylinders have a
galaxy in their centre and are oriented along the LOS passing for that
the centre point or orthogonally to it. In the former case we estimate
$\langle n^p(r;h) \rangle$ and in the latter $\langle n^o(r;h)
\rangle$.  If there are no selection effects, { redshift space
  distortion are negligible} and the distribution is isotropic we
expect $\langle n^p(r;h) \rangle \approx \langle n^o(r;h) \rangle
\approx \langle n(r) \rangle$. Any radial dependent effect, such as
spurious observational selection effects, intrinsic physical effects
depending on distance (e.g., galaxy evolution, etc.) will alter this
situation.

 {We have then measured the conditional density with the FS estimator
   { in spherical volumes} in several redshift surveys, i.e. the
   main galaxy sample of SDSS, the LRG-SDSS samples, the QSO-SDSS
   samples, the 2dFGRS samples and the 2MRS samples. { We find in
     all samples that at small scales, i.e. $r \le 20$ Mpc/h the
     exponent is $\gamma \approx 0.9$, while on large scales, in the
     SDSS VL samples which cover larger volumes, there is evidence of
     a change of slope, where we estimate $\gamma \approx 0.3$ in the
     range $20-100$ Mpc/h}. }
{ For the case of the LRG sample we concluded further work is required
to determine whether selection effects introduce a relevant bias in
the measurement of its statistical properties. Indeed, although this
shows a relatively small selection, i.e. $\delta^p(r) \approx 0.05$,
this rapidly grows toward its boundary (i.e., $r\in [100,200]$ Mpc/h).}

We have then used the determination of the conditional density by
means of the FS cylinder method to constraint large scale selection
effects in the data. In particular, we have introduced a new method,
the gradient cylinder method (CGM), which is able to detect radial
dependent selection effects in three dimensional galaxy samples {
  without the assumption of spatial homogeneity of the underlying
  distribution. }
In practice, we define a normalised quantity along the LOS,
$\delta^p(r;h)$, or perpendicularly to it $\delta^o(r;h)$, and we measure
its amplitude and its dependence of the distance $r$.  We have tested
it in artificial samples with a priori known proprieties. We have
shown, by considering artificial point distribution with known
properties, that the CGM is able to correctly identify the selection
effects that were introduced in the data. In particular, when
$\delta^p(r;h) > 0.1$ selection effects maybe strong enough to influence
the behaviour of the conditional density measured with FS estimators in
spheres.  On the other hand we have also found that the conditional
density is insensitive to moderate selection effects in the data.

We have applied the CGM method to the above mentioned redshift
surveys.  The main results is that almost in all samples we do not
find a large { radial dependence}. The samples that are mostly
influenced by these effects are the deeper VL samples of the SDSS and
the LRG samples, where, at large scales $\delta^p(r;h) \approx 0.1-0.2$.
Instead for the other VL samples of SDSS, for the 2dFGRS and the 2MRS
samples we find $\delta^p(r;h) \approx \delta^o(r;h) \le 0.1$. On the
other hand, the QSO-DR7 samples are completely dominated by radial
dependent effects (physical QSO evolution and luminosity selection
effects) that it is not possible to make a reliable correlation
analysis in these samples.

 { In summary we have introduced a new method to study large scales
   correlation properties of galaxy distribution and that allows to
   control redshift dependent selection effects.  The application of
   these methods to forthcoming redshift surveys, like the extension
   of the SDSS \cite{boss}, when they will cover a contiguous and
   large sky area, will be able to clarify the nature of galaxy
   correlations on scales $> 100$ Mpc/h.}

%%%%%%%%%%%%%%%%%%%%%%%%%%%%%%%%%%%%%%%%%%%%%%%%%%%%%%%%%%%%%

\acknowledgments We acknowledge Andrea Gabrielli and Micheal Joyce for
useful comments and discussion.  
DT and YB thanks for the partial financial support the
Saint-Petersburg State University research projects No.6.38.669.2013
and No.6.38.18.2014. 
DT thanks the Institute for Complex System of CNR
for the kind hospitality during the writing of this paper.

We acknowledge the use of the 2dFGRS data \footnote{{\tt
    http://www.mso.anu.edu.au/2dFGRS/ }}, of the 2MRS
data \footnote{{\tt
    https://www.cfa.harvard.edu/$\sim$dfabricant/huchra/2mass/}},and
of the millennium run semi-analytic galaxy catalogues \footnote{{\tt
    http://www.mpa-garching.mpg.de/galform/agnpaper/}}.  SDSS-III is
managed by the Astrophysical Research Consortium for the Participating
Institutions of the SDSS-III Collaboration including the University of
Arizona, the Brazilian Participation Group, Brookhaven National
Laboratory, University of Cambridge, Carnegie Mellon University,
University of Florida, the French Participation Group, the German
Participation Group, Harvard University, the Instituto de Astrofisica
de Canarias, the Michigan State/Notre Dame/JINA Participation Group,
Johns Hopkins University, Lawrence Berkeley National Laboratory, Max
Planck Institute for Astrophysics, Max Planck Institute for
Extraterrestrial Physics, New Mexico State University, New York
University, Ohio State University, Pennsylvania State University,
University of Portsmouth, Princeton University, the Spanish
Participation Group, University of Tokyo, University of Utah,
Vanderbilt University, University of Virginia, University of
Washington, and Yale University.

\appendix

\section{The data}
\label{data} 

 { We briefly summarise the main properties of the redshift surveys for
   which we have analysed correlation properties.

\subsection{SDSS}
 
The SDSS \citep{york2000} is currently the largest spectroscopic
survey of extragalactic objects containing redshifts for more than
1,000,000 galaxies and 100,000 quasars.  There are two independent
parts of the galaxy survey in the SDSS: the main galaxy (MG) sample
and the luminous red galaxy (LRG) sample.

\subsubsection{The main galaxy sample} 
\label{mg-sample} 
For the MG sample  we have considered
spectroscopic catalogue SDSS-DR7 (see \cite{paper_dr6} and the SDSS web
site {\tt http://www.sdss.org}).
The spectroscopic survey covers an area of about 10,000 square degrees
on the  celestial sphere. The Petrosian apparent  magnitude limit with
extinction corrections for the galaxies is 17.77 in the $r$-filter and
photometry for  each galaxy  is available in  five different  bands. A
detailed discussion of the  spectroscopic target selection in the SDSS
MG sample can be found in \cite{strauss2002}.

We have constructed several volume limited (VL) samples that are
unbiased for the selection effect related to the cuts in the apparent
magnitude. To this aim we have applied the standard procedure
described, for instance in \cite{zehavietal05} by considering the
following steps (we refer the interested reader to \cite{sdss_aea} for
more details).

\begin{itemize} 
\item We selected only the galaxies from the MG
sample.  

\item We considered galaxies in the redshift range $10^{-4} \leq
z \leq 0.3$ with redshift confidence $z_{conf} \ge 0.35$.
\item We applied the apparent magnitude filtering condition $m_r <
17.77$  \cite{strauss2002}.  
\item We  considered galaxies in the angular region 
limited, in the SDSS internal angular coordinates, by $-33.5^{\circ}
\le \eta \le 36.0^\circ$ and $-48.0^\circ \le \lambda \le 51.5^\circ$.
\item  We do not use
corrections for the redshift completeness mask or for fiber collision
effects. Both are estimated to be small. 
\item We computed the metric distances $r(z) = {c \over H_0}
  \int_{1/(1+z)}^1 {1 \over y \sqrt{0.3/y + 0.7 y^2}} $ using the
  standard cosmological parameters, {\it i.e.,} $\Omega_M=0.3$ and
  $\Omega_\Lambda=0.7$. Results are given in units of $h$ defined as
  $H_0=$100 h km/sec/Mpc\footnote{These same values of the
    cosmological parameters are chosen also for the other surveys
    discussed in what follows}.
\item We computed absolute
magnitudes $M_r$ using Petrosian apparent magnitudes in the $m_r$
filter corrected for Galactic absorption and  we used standard
K-correction from the VAGC data \footnote{{\tt
    http://sdss.physics.nyu.edu/vagc/}}.
\item Details of the VL samples are reported in Tab.\ref{tab_sdss}. 
\end{itemize}

\begin{table}
\begin{center}
\begin{tabular}{|c|c|c|c|c|c|}
  \hline
  VL sample & $R_{min}$ & $R_{max}$ & $M_{min}$  & $M_{max}$ & N\\
  \hline
    VL1    & 50  & 200 & -18.9 & -21.1   &73810\\
    VL2    & 100 & 300 & -19.9 & -22.0   &110570\\
    VL3    & 125 & 400 & -20.5 & -22.2   &129974\\
    VL4    & 150 & 500 & -21.1 & -22.4   &94179\\
    VL5    & 200 & 600 & -21.6 & -22.8   &51697\\
    VL6    & 70  & 450 & -20.8 & -21.8   &112860\\
   \hline
\end{tabular}
\end{center}
\caption{Main properties of the obtained SDSS VL samples: $R_{min}$,
  $R_{max}$ (in Mpc/h) are the chosen limits for the metric distance;
  ${M_{min}, \,M_{max}}$ define the interval for the absolute
  magnitude in each sample and $N$ is the number of galaxies included
  in the sample.}
\label{tab_sdss}
\end{table}

%%%%%%%%%%%%%%%%%%%%%%%%%%%%%%%%%%%%%%%%%%%%%
\subsubsection{The luminous red galaxy sample} 
\label{lrg-sample} 

 {The selection of the LRG galaxies is discussed in \cite{lrg} while
   the construction of the different sub-samples considered here is
   described in detail by \cite{kazin}. Briefly, we focus on the
   so-called DR7-Dim which is limited by $0.16<z<0.36$. Indeed, for
   $z>0.36$ there is a clear selection effect due to the passage of
   the 4000 A break into the $r$ band \cite{kazin}.  This corresponds
   to the sharp decrease of the redshift counts for $z>0.36$.  }

 {The limits for the DR7-Dim sample are: $R_{min} = 465 $ Mpc/h and
  $R_{max} = 1002$ Mpc/h. The limits in R.A. $\alpha$ and
  Dec. $\delta$ considered are chosen in such a way that (i) the
  angular region does not overlap with the irregular edges of the
  survey mask and (ii) the sample covers a contiguous sky area.  Thus
  we have chosen: $\alpha_{min} = 130^\circ \le \alpha \le
  \alpha_{max} = 240^\circ$; $\delta_{min} = 0^\circ \le \delta \le
  \delta_{max} = 50^\circ$ . The absolute magnitude is constrained in
  the range $M \in [-23.2,-21.2]$. With these limits we find
  $N=41,833$ galaxies covering a solid angle $\Omega=1.471$ sr.  }

 {As we are interested in determining
the role of selection effects, we have also considered the LRG sample
SDSS-Full limited at $z<0.4$, containing $N=65,470$ galaxies.}

%%%%%%%%%%%%%%%%%%%%%%%%%%%%%%%%%%%%%%%%%%%%%%%

\subsubsection{The quasar sample} 
\label{qso-sample} 

 { We have considered fifth edition of the SDSS quasar (QSO) catalogue
   described in \cite{Schneider2010} whose main part consists of a
   contiguous area of about 7,600 deg$^2$ in the North Galactic
   Pole. The details of the construction of the QSO catalogue are
   presented in \cite{Schneider2010} (and references therein). In this
   paper we focus on a same sub-sample analysed by
   \cite{clowes2013,nadathur2013} 
whose main characteristics are: 
\begin{itemize} 
\item  a redshift range  $z \in[1.0,1.8]$
\item an  angular region  limited
   by $130 ^\circ \le \alpha \le 235^\circ$ and $0^\circ \le \delta
   \le 60^\circ$
\item The $i-$band apparent magnitude limit is $i \le
   19.0$. 
\end{itemize} 
The sample constructed in this way is ML and contains
   N=18,722 QSO. We have then constructed a VL sample by imposing
   the additional limits in distance, $R_{max} \le 4545$ Mpc/h, and
   absolute magnitude, $M_{min}=-26.0$ and selecting in this way
   N=6351 objects.

\subsection{The Two-degree Field Galaxy Redshift Survey}

\label{2dfgrs} 

The Two-degree Field Galaxy Redshift Survey (2dFGRS) \cite{colless01}
measured redshifts for more than $220,000$ galaxies in two strips, one
in the Southern Galactic Cap (SGC) and the other in the Northern
Galactic Cap (NGC).  As for the SDSS survey, we have constructed some
VL samples (more details can be found in \cite{2df_aea}):

\begin{itemize} 

\item The apparent magnitude corrected for galactic extinction in the
  $b_J$ filter is limited to $14.0<b_J<19.45$.

\item We selected two rectangular regions: in the SGC there is a slice
  of size $84^\circ\times 9^\circ$ limited by
  $-33^\circ<\delta<-24^\circ$, $-32^\circ<\alpha<52^\circ$, while the
  NGC slice is smaller, i.e., $60^\circ\times 6^\circ$, with limits
  $-4^\circ<\delta<2^\circ$, $150^\circ<\alpha<210^\circ$ 

\item Galaxies have redshifts in the range  $0.01 \leq z \leq 0.3$. 

\item We did not use a correction for the redshift-completeness mask
  and for the fiber collision effects,  { which are negligible in the
  region we considered.} 

\item We applied K-correction $K(z)$ as in \cite{2df_aea}.
 
\item Details of the VL samples are reported in 
  Table~\ref{tbl_VLSamplesProperties}.
\end{itemize}

\begin{table}
\begin{center}
\begin{tabular}{|l|c|c|c|c|c|}
  \hline
  VL sample & $R_{min}$ & $R_{max}$ & $M_{min}$ & $M_{max}$ & $N_g$\\
  \hline
    SGC400 & 100 & 400 & -20.8 & -19.0 & 29373 \\
    NGC400 & 100 & 400 & -20.8 & -19.0 & 23208 \\
    SGC550 & 150 & 550 & -21.2 & -19.8 & 26289 \\
    NGC550 & 150 & 550 & -21.2 & -19.8 & 18030 \\
  \hline
\end{tabular}
\end{center}
\caption{Main properties of the obtained VL samples.  $R_{min}$,
  $R_{max}$ are the chosen limits for the metric distance; ${M_{min},
    \,M_{max}}$ are the corresponding limits in the absolute
  magnitude; $N_g$ is the number of galaxies in the sample. }
\label{tbl_VLSamplesProperties}
\end{table}

\subsection{The Two Micron All Sky Galaxy Redshift Survey} 

\label{2mass} 

The Two Micron All Sky Galaxy Redshift Survey (2MRS) contains sample
of 44599 galaxies with $K_s < 11.75$ mag and $|b| > 5^\circ$ ($>
8^\circ$ towards the Galactic Centre) \cite{Huchra2012}.  This sample
of the near-infrared all-sky galaxies contains 44599 homogeneously
selected spectral galaxy redshifts. From the ML sample we construct 8
(VL) samples (see Tab.\ref{tab_2mrs}) 4 in the Northern (N) and 4 in
the Southern hemispheres. No additional correction have been used
because the sample is at low-redshift and because K-corrections are
small in the near infrared.
\begin{table}
\begin{center}
\label{tab:VL}
\begin{tabular}{|l|c|c|c|c|}
\hline
~& 	$R_{max}$ & $M_K$ & N & $ \Lambda$ \\
\hline
VL1N & 50  & -21.87 & 1845 & 1.51\\ 
VL2N & 100 & -23.39 & 4768 & 2.27\\ 
VL3N & 150 & -24.27 & 6133 & 3.26\\ 
VL4N & 200 & -24.95 & 3741 & 5.64\\ 
\hline                        
VL1S & 50  & -21.87 &  936 & 1.93\\ 
VL2S & 100 & -23.39 & 4925 & 2.23\\ 
VL3S & 150 & -24.27 & 5923 & 3.28\\ 
VL4S & 200 & -24.95 & 4015 & 5.38\\ 
\hline
\end{tabular}
\caption{Parameters of the considered VL samples of 2MRS galaxies:
  $R_{max}$ limiting of the considered VL samples of 2MRS galaxies:
  $M_K$ limiting absolute magnitude in the $K_s$ filter, $N$ - number
  of objects and $\Lambda$ is the mean distance between nearest
  neighbours in $Mpc/h$.} 
\label{tab_2mrs}
\end{center}
\end{table}

}

%%%%%%%%%%%%%%%%%%%%%%%%%%%%%%%%%%%%%%%%%%%%%%%%%%%%%%%%%%%%%%%%5

\end{document}